\documentclass[draft]{article}
\usepackage{amsmath,amsfonts,latexsym, amssymb}

\usepackage{color}

\newcommand{\ov}{\overline}
\newcommand{\const}{\mbox{const}}

\newcommand{\e}{\varepsilon}
\newcommand{\eps}{\varepsilon}

\newcommand{\rd}{{\rm d}}
\newcommand{\bR}{{\mathbb R}}

\newcommand{\bs}{{\bf{s}}}
\newcommand{\ba}{{\bf{a}}}
\newcommand{\bb}{{\bf{b}}}
\newcommand{\bx}{{\bf{x}}}

\newcommand{\bt}{{\bf{t}}}
\newcommand{\bu}{{\bf{u}}}
\newcommand{\bv}{{\bf{v}}}

\newcommand{\bw}{{\bf{w}}}
\newcommand{\bz}{{\bf{z}}}

\newcommand{\wh}{\widehat}

\newcommand{\al}{\alpha}

\newcommand{\be}{\begin{equation}}
\newcommand{\ee}{\end{equation}}

\newcommand{\om}{{\omega}}

\newcommand{\cA}{{\cal A}}

\newcommand{\cN}{{\cal N}}
\newcommand{\cP}{{\cal P}}

\newcommand{\bP}{{\bf P}}
\newcommand{\bQ}{{\bf Q}}

\newcommand{\im}{{\text{Im} }}
\newcommand{\re}{{\text{Re} }}

\newcommand{\E}{{\mathbb E }}

\newcommand{\N}{{\mathbb N}}
\renewcommand{\P}{{\mathbb P}}
\newcommand{\bC}{{\mathbb C}}

\newcommand{\wt}{\widetilde}

\newtheorem{theorem}{Theorem}
\newtheorem{corollary}[theorem]{Corollary}
\newtheorem{lemma}[theorem]{Lemma}
\newtheorem{proposition}[theorem]{Proposition}

\newcommand{\qed}{\hfill\fbox{}\par\vspace{0.3mm}}

\numberwithin{equation}{section}
\numberwithin{theorem}{section}
\numberwithin{definition}{section}
\title{Wegner estimate and level repulsion for Wigner random matrices}

\author{L\'aszl\'o Erd\H os${}^1$\thanks{Partially supported
by SFB-TR 12 Grant of the German Research Council}, Benjamin
Schlein${}^2$\thanks{Partially supported by Sofja-Kovalevskaya Award of
the Humboldt Foundation.}\;
and Horng-Tzer Yau${}^3$\thanks{Partially supported
by NSF grants DMS-0602038, 0757425, 0804279} \\
\\
Institute of Mathematics, University of Munich, \\
Theresienstr. 39, D-80333 Munich, Germany${}^1$ \\ \\
Department of Pure Mathematics and Mathematical Statistics
\\  University of Cambridge \\
Wilberforce Rd, Cambridge CB3 0WB, UK${}^2$ \\ \\
Department of Mathematics, Harvard University\\
Cambridge MA 02138, USA${}^3$ \\ \\
\\}

\begin{document}
\date{May 3, 2009}

\maketitle

\begin{abstract}

We consider  $N\times N$ Hermitian random matrices with
independent identically distributed
entries (Wigner matrices).  The matrices are normalized so that
the average spacing between consecutive eigenvalues is of order $1/N$.
Under suitable assumptions on the distribution
of the single matrix element, we first prove that,  away from the spectral
edges, the empirical
density of eigenvalues concentrates around the Wigner semicircle law
on energy scales $\eta \gg N^{-1}$. This result establishes the semicircle law
on the optimal scale and it removes a logarithmic factor
from our previous result \cite{ESY2}. We then show a
Wegner estimate, i.e.  that the averaged
density of states is bounded.
Finally, we prove that the eigenvalues of
a Wigner matrix repel each other, in agreement with the
universality conjecture.

\end{abstract}


\medskip

{\it Running title:} Wegner estimate and level repulsion

\medskip


%\received{}

\bigskip

\section{Introduction}

Let $H=(h_{ij})$ be an $N\times N$ hermitian matrix, $N\ge 2$,
and let
$\mu_1\leq \mu_2 \leq\ldots \leq \mu_N$ denote its eigenvalues.
These matrices form a {\it hermitian Wigner ensemble} if the
matrix elements,
\be
h_{i j} = \bar{h}_{ji}=  N^{-1/2}z_{ij}\in \bC,
\quad (1\leq i < j\leq N), \quad \text{and}
\quad   h_{i i} =  N^{-1/2}  x_{ii}\in \bR, \quad (1\leq i\leq N)
\label{wig}
\ee
are independent random variables with  mean zero.
We assume that $z_{ij}$ ($i< j$)
all have a common distribution $\nu$ with variance
$\int_\bC |z|^2 \rd\nu(z)=1$
and with a strictly positive density function $h:\bR^2\to \bR_+$, i.e.
$$
\rd \nu(z)= \mbox{(const.)}h(x,y)\rd x\rd y\quad \mbox{where}\quad
x=\re \, z,\;\;\
y=\im \, z.
$$
We will often denote $g:= -\log h$. Throughout the paper
we also assume that
\be
   \mbox{either} \quad h(x,y)=  h^*(x)h^*(y), \qquad
 \mbox{or} \quad  h(x,y)=  h^* (x^2+y^2)
\label{hass}
\ee
with some positive function $h^*:\bR\to\bR_+$,
i.e. either the real and imaginary parts of the random variables $z_{ij}$,
$i<j$, are independent and identically distributed, or the distribution
depends only on the absolute value $|z_{ij}|$.
The diagonal elements, $x_{ii}$,  also have  a
common distribution, $\rd\wt \nu(x)=\mbox{(const.)}e^{-\wt g(x)}\rd x$
with $\wt g: \bR\to \bR$.
Let  $\P$ and $\E$
denote the probability and the expectation value, respectively,
w.r.t the joint distribution of all matrix elements.
The normalization \eqref{wig}
of the matrix elements and fixing the variance of
$\rd\nu$ to be one
ensure that the spectrum of $H$ is $[-2,2]+o(1)$
with probability one in the limit as $N\to \infty$.

For the special case $g(x,y)=x^2+y^2$, $\wt g(x)=x^2/2$,
the hermitian Wigner ensemble is called the {\it Gaussian
Unitary Ensemble (GUE)}. Due to the unitary invariance of the GUE
matrices, the joint eigenvalue distribution can be explicitly
expressed in terms of a Vandermonde determinant and all correlation
functions are computable  (see \cite{Mehta} for an overview).
This approach can be applied for more general ensembles with
unitary invariance, i.e. for ensembles where the distribution
is invariant under the transformation $H\to U^{-1}HU$ for any
unitary matrix $U$
(for a general overview via the Riemann-Hilbert
approach see \cite{D}).
In particular, the density of the eigenvalues converges
to the Wigner semicircle law as $N\to \infty$ and the truncated
two-point correlation function, appropriately rescaled,
is given by the famous Wigner-Dyson sine-kernel in the bulk spectrum,
see \cite{D, PS} and references therein
(near the spectral edges a  different universal statistics holds).
Higher order correlations can be expressed as determinants
involving the sine-kernel. The order statistics of eigenvalues can
also be  computed. The most important one is the nearest-neighbor
level statistics, or gap distribution,
i.e. the distribution of the difference between
two consecutive eigenvalues, $\mu_{\al+1}-\mu_\al$, in the bulk. With an
appropriate rescaling, the density function $f(x)$ of the eigenvalue gap
is universal. It is characterized by $f(x)\sim x^2$
near 0 that corresponds to a strong level repulsion. The large
distance behavior, $f(x)\sim \exp(-x^2)$, $x\gg 1$,
expresses a strong supression
of large eigenvalue gaps.

These properties of the eigenvalue statistics are conjectured to hold
for much more general matrix ensembles beyond the invariant ensembles,
in particular
for general Wigner matrices.
Numerical evidences very strongly support these conjectures,
nevertheless only a  few rigorous results are known
for ensembles without unitary invariance (notable exceptions
are the universality of the Tracy-Widom distribution for the
extremal eigenvalues \cite{Sosh} and the Wigner-Dyson sine-kernel
for Wigner matrices with Gaussian convoluted distributions \cite{J}).
In this paper we prove the strong level repulsion and a subexponential
estimate for the large distance behavior of the gap distribution.

\bigskip

For Wigner matrices,
the Wigner semicircle law has  been long established on scale of order 1,
i.e. the empirical counting measure of the
eigenvalues (also called empirical density of states measure
in physics), $\varrho_N(E)=\frac{1}{N}\sum_{\al=1}^N
\delta(E-\mu_\al)$, converges weakly to $\varrho_{sc}(E)\rd E$ (see
\eqref{scform}) in probability as $N\to\infty$ (see \cite{W}
for the original result).
The weak convergence does not
allow one to identify the local density of eigenvalues on energy
scales $\eta \ll 1$.
Note that the number of eigenvalues
in any interval of length $\eta$ within $[-2,2]$ is typically  of order
$N\eta$, so the self-averaging property
is expected to hold for the smoothed density of states as long as
the smoothing is on  scale $\eta\gg 1/N$.
In Section \ref{sec:sc}, using
a necessary a-priori bound from Section \ref{sec:upp}, we prove that
the semicircle law holds on the smallest possible scales, i.e.
for any interval of length $\eta\gg 1/N$ (Theorem \ref{thm:sc1}).
This removes the
logarithmic factor in our previous work \cite{ESY2} and establishes
the optimal result.
As a corollary, we obtain an optimal result on the delocalization
of the eigenvectors (Corollary \ref{cor:linfty}).
The proof is a bootstrap argument in $\eta$; it
relies on (non-optimal) bounds on the supremum norm of the eigenvectors, which
in turn, can be obtained by first
establishing the semicircle law on a larger scale
$\eta\ge (\log N)^4/N$. Although the semicircle law on
a larger scale and  bounds on the eigenvectors
were already established in \cite{ESY2},
the error bound was not sufficiently strong. Therefore,
in Section \ref{sec:larger}, we first improve the results
of \cite{ESY2}.

In Section \ref{sec:tail} we give an upper bound on the tail
distribution of the distance between consecutive eigenvalues
(Theorem \ref{thm:gap}).
The bound is only subexponential in contrast to the expected
Gaussian decay. In Section \ref{sec:wegner} we prove the Wegner
estimate for Wigner matrices, i.e. that the averaged density of states,
$\E\, \varrho_N(E)$, is uniformly bounded (Theorem \ref{thm:wegner}).
Note that the Wegner estimate is an information on
arbitrarily short scales, i.e. it is  uniform in $\eta$.
On scales $\eta \lesssim 1/N$, however, the smoothed empirical
density of states truly fluctuates since
individual eigenvalues near $E$ dominate, but the averaged
density of states remains bounded.

Finally, in Section \ref{sec:level} we establish
an upper bound $f(x)\leq Cx^2$ for the density function
of the eigenvalue spacing in the regime where $x$ is small
(Theorem \ref{thm:repul}).
Apart from the constant,
this  upper bound coincides with the prediction obtained
from the universality conjecture on the level spacing distribution
and it proves that the level repulsion in Wigner
matrices is as strong as in the GUE ensemble.
We also give an optimal estimate
on higher order level repulsion. We show that the probability
that there are $k$ eigenvalues in a small spectral interval $I$, with
$|I|=\e/N$ ($\e\ll 1$), is bounded from above by $C\e^{k^2}$ in accordance
with the prediction from GUE that is based upon the explicit formula
for the joint density function $\sim \prod_{j<\ell}(\mu_\ell-\mu_j)^2$
of $k$ eigenvalues.

\medskip

We work with hermitian Wigner matrices, but our method applies to
symmetric Wigner matrices as well. In that case, the level repulsion
is weaker, $f(x)\leq Cx$, in accordance with the explicit
gap distribution function for Gaussian Orthogonal Ensemble (GOE).

\bigskip

We need to assume further conditions
on the distributions
of the matrix elements in addition to \eqref{wig}, \eqref{hass}:
\begin{itemize}
\item[{\bf C1)}] There exists a $\delta_0>0$ such that
\be
D:=\int_\bC \exp \big[ \delta_0 |z|^2\big]
\rd \nu(z)  <\infty , \qquad
\wt D:=\int_\bR \exp{\big[\delta_0 x^2\big]}\rd \wt\nu(x)  <\infty \; .
\label{x2}
\ee
\end{itemize}
To establish the Wegner estimate and the level repulsion,
we need some smoothness property
of the density function $h$. We assume that
\begin{itemize}
\item[{\bf C2)}] The Fourier transform
of the functions $h$ and
$h(\Delta g)$, with $g=-\log h$,
satisfies the decay estimate
\be
 |\wh h(t,s)|\leq \frac{1}{\left[1+\om_a(t^2+s^2)\right]^a},
\qquad |\wh{h\Delta g}(t,s)|
\leq \frac{1}{\left[1+\wt\om_a(t^2+s^2)\right]^a}
\label{charfn}
\ee
with some exponent $a\ge 1$ and constants $\om_a, \wt\om_a> 0$.
(Note that $a\om_a\leq \frac{1}{4}$ by the condition
that the variance is 1.)
\end{itemize}

\medskip

\noindent
In our previous papers \cite{ESY, ESY2} we
assumed that $\rd\nu$ satisfies the logarithmic Sobolev inequality
for the proof of the analogue of Lemma \ref{lm:x-old}
(Lemma 2.1 of \cite{ESY2}).
M. Ledoux has kindly pointed out to us that by
applying a theorem of Hanson and Wright \cite{HW},
this lemma also holds under
the moment condition  {\bf C1)} only.
We remark that the original paper \cite{HW}
assumed that $\rd\nu$ was symmetric; this conditon was
later removed by Wright \cite{Wr}.

Another assumption we made in \cite{ESY, ESY2} states that
either  the Hessian of $g=-\log\, h$ is bounded from above
or the distribution is compactly supported. This was needed
because we used  Lemma 2.3 of \cite{ESY}, whose original
proof required the condition on $\mbox{Hess}\; g$.
An alternative proof of this lemma was given by Bourgain
(the proof reproduced in the Appendix of \cite{ESY2}) under the
additional condition that the support of $\rd \nu$ is compact.
In this paper, we extend the results of  \cite{HW, Wr}
and apply them to prove a weaker but for our purposes
still sufficient version of  Lemma 2.3 in \cite{ESY}.
This approach requires no additional condition apart
from  {\bf C1)}.
Condition ${\bf C2)}$ will play a role only in
Theorem \ref{thm:wegner} and Theorem \ref{thm:repul}.

In our previous papers \cite{ESY, ESY2} we
assumed that the real and imaginary parts of $z_{ij}$
are independent. It is straightforward to check that all results
of \cite{ESY, ESY2} hold for the case of radially symmetric
distributions (second condition in \eqref{hass})
as well.

\medskip

{\it Convention.} We assume  condition {\bf C1)} throughout the paper
and every constant  may depend on the constants $\delta_0, D, \wt D$
from \eqref{x2} without further notice.

\bigskip

{\it Acknowledgement.} The authors are grateful to M. Ledoux
for his remark that  Lemma 2.1 of \cite{ESY2} follows
from a result of Hanson and Wright \cite{HW}.

\section{Notation and the basic formula}\label{sec:not}

For any spectral
parameter $z= E+i\eta\in \bC$, $\eta>0$, we denote the Green function
by $G_z= (H-z)^{-1}$. Let $F(E)=F_N(E)$ be
the empirical distribution function of the eigenvalues
\be
F(E):= F_N(E)=
\frac{1}{N}\big| \, \big\{ \al \; : \; \mu_\al \leq E\big\}\Big|\;
\label{Fdef}
\ee
(in physics it is called the {\it integrated density of states}).
Its derivative is the {\it empirical density of states} measure
$$
 \varrho(E): =  F'(E) = \frac{1}{N}\sum_{\al=1}^N \delta(E-\mu_\al) .
$$
Its  statistical average, $\E \, \varrho(E)$,
is called the {\it averaged density of states.}
We define the Stieltjes transform of $F$ as
\be
 m= m(z) =\frac{1}{N}\text{Tr} \; G_z = \int_\bR \frac{\rd F(E)}{E-z}\,,
\label{Sti}
\ee
and we let
\be
\varrho_{\eta}(E) = \frac{ \text{Im} \;  m(z)}{\pi}=
\frac{1}{N\pi} \text{Im} \; \text{Tr} \; G_z
=\frac{1}{N\pi}\sum_{\al=1}^N \frac{\eta}{(\mu_\al-E)^2+\eta^2}
\label{rhodef}
\ee
be the normalized density of states of $H$
around energy $E$ and regularized on scale $\eta$.
We note that
$\varrho(E) = \lim_{\eta\to 0+0}\varrho_\eta(E)$.
The random variable $m$ and the random measures
$\varrho$ and $\varrho_\eta$  also depend
on $N$, when necessary, we will indicate this fact  by
writing $m_N$, $\varrho_N$ and $\varrho_{\eta,N}$.

For any $z=E+i\eta$, $\eta\ne 0$, we let
$$
m_{sc}= m_{sc}(z) = \int_\bR \frac{\varrho_{sc}(x)\rd x}{x - z}
$$
be the Stieltjes transform of the Wigner semicircle
distribution function whose density is given by
\be
\varrho_{sc}(E) = \frac{1}{2\pi} \sqrt{4-E^2}
 {\bf 1}(|E|\leq 2)\; .
\label{scform}
\ee

\bigskip

Let $B^{(k)}$ denote the $(N-1)\times(N-1)$ minor of $H$ after
removing the $k$-th row and $k$-th column. Note that
$B^{(k)}$ is an $(N-1)\times(N-1)$ Hermitian Wigner matrix
with a normalization factor off by $(1-\frac{1}{N})^{1/2}$.
Let
$\lambda_1^{(k)}\leq \lambda_2^{(k)}\leq \ldots \leq \lambda_{N-1}^{(k)}$
denote its eigenvalues and $\bu_1^{(k)},\ldots , \bu_{N-1}^{(k)}$
the corresponding normalized eigenvectors.

Let $\ba^{(k)}=(h_{k,1}, h_{k,2}, \ldots h_{k,k-1}, h_{k,k+1}, \ldots
h_{k,N})^*
\in \bC^{N-1}$, i.e. the $k$-th column after removing the diagonal
element $h_{k,k}=h_{kk}$. Computing the $(k,k)$ diagonal element of
the resolvent $G_z$, we have
\be
 G_z(k,k)= \frac{1}{h_{kk}-z-\ba^{(k)}\cdot (B^{(k)}-z)^{-1}\ba^{(k)}}
 = \Big[ h_{kk}-z-\frac{1}{N}\sum_{\alpha=1}^{N-1}\frac{\xi_\al^{(k)}}
{\lambda_\al^{(k)}-z}\Big]^{-1},
\label{mm}
\ee
where we defined
$$
  \xi_\al^{(k)} : = \big| \sqrt{N}\ba^{(k)}\cdot \bu_\al^{(k)}\big|^2
$$
and note that $\E \, \xi_\al^{(k)}=1$.
Thus
\be
m(z)
=\frac{1}{N}\sum_{k=1}^N \Bigg[ h_{kk} - z -
\frac{1}{N}\sum_{\al=1}^{N-1}
\frac{\xi^{(k)}_\al}{\lambda_\al^{(k)}-z}\Bigg]^{-1}\;.
\label{mm1}
\ee
Similarly to the definition of $m(z)$ in \eqref{Sti},
we also define the Stieltjes transform
of the density of states of $B^{(k)}$
$$
m^{(k)}= m^{(k)}(z) = \frac{1}{N-1}\, \text{Tr}\, \frac{1}{B^{(k)}-z}
 =\int_\bR \frac{\rd F^{(k)}(x)}{x - z}
$$
with the empirical counting function
$$
  F^{(k)}(x) = \frac{1}{N-1} \big| \, \big\{ \al \; : \;
 \lambda_{\al}^{(k)}\leq x \big\}\big|.
$$
The spectral parameter $z$ is fixed in most of the proofs
and we will often omit it from the argument of the Stieltjes transforms.
Let $\E_k$ denote the expectation value w.r.t
the random vector $\ba^{(k)}$. The distribution of $B^{(k)}$, $\ba^{(k)}$
and $\xi_\al^{(k)}$ does not depend on $k$, so we will often
omit this superscript when it is unnecessary.

For any spectral interval $I\subset \bR$, we denote
$$
  \cN_I: =\#\{ \alpha\; :\; \mu_\al\in I\}
$$
$$
  \cN_I^{(k)}:= \#\{ \alpha\; : \; \lambda_\al^{(k)}\in I\}
$$
the number of eigenvalues in $I$ of $H$ and $B^{(k)}$, respectively.
When we are interested only in the distribution of $\cN_I^{(k)}$,
we drop the superscript $k$, but to avoid confusion with $\cN_I$,
we denote by $\cN^\lambda_I$ a random variable with the common
distribution of $\cN_I^{(k)}$.

\bigskip

With these notations, the following basic
upper bound on $\cN_I$ follows immediately:

\begin{proposition}\label{prop:basic}
Let $I=[E-\eta/2, E+\eta/2]$ be an interval  of length $\eta>0$ about
the spectral point $E\in \bR$ and let $z=E+i\eta$.
Then we have the following estimate on the number of eigenvalues in $I$:
\be
\begin{split}
 \cN_I \leq &\;  C\eta \, \im \sum_{k=1}^N \Bigg[ h_{kk} - z -
\frac{1}{N}\sum_{\al=1}^{N-1}
\frac{\xi^{(k)}_\al}{\lambda_\al^{(k)}-z}\Bigg]^{-1}\; .
\label{basic}
\end{split}
\ee
\end{proposition}

{\it Proof.}  We have
$$
 \cN_I = N\int_I \rd F(x) \leq \frac{5}{4}N\eta\int_{E-\eta/2}^{E+\eta/2}
 \frac{\eta\rd F(x)}{(x-E)^2+\eta^2} \leq \frac{5}{4}N\eta \, \im \, m(z)
$$
and using \eqref{mm1} we obtain \eqref{basic}. \qed

\section{Main results}

The first main result establishes the semicircle law  on the optimal
scale $\eta\ge O(1/N)$; the proof will be given in Section \ref{sec:sc}.

\begin{theorem}\label{thm:sc1}
Let $H$ be an $N\times N$ hermitian Wigner matrix satisfying the
condition {\bf C1)}.
Let $\kappa>0$ and
fix an energy $E\in [-2+\kappa, 2-\kappa]$.
Then there exist positive constants $C$, $c$, depending
only on $\kappa$, and a universal constant $c_1>0$ such that the following hold:
\begin{itemize}
\item[(i)] For any $\delta \le c_1 \kappa$  and $N\ge 2$ we have
\be
 \P (  |m(E+i\eta)-m_{sc}(E+i\eta)|\ge \delta)
\leq C\, e^{-c\delta\sqrt{N\eta}}
\label{sc:new}
\ee
for any $K/N \leq \eta \le 1$, where $K = 300/c_0$ and
$c_0:=\pi \varrho_{sc}(E)>0$.

\item[(ii)] Let
$\cN_{\eta^*}(E)= \cN_{I^*}$
denote the number of eigenvalues in
the interval $I^*:=[E-\eta^*/2, E+\eta^*/2]$. Then
for any $\delta \leq c_1 \kappa$ there is a constant $K_\delta$, depending only on
$\delta$, such that
\be
\P \Big\{  \Big| \frac{\cN_{\eta^*}(E)}{N\eta^*}
- \varrho_{sc}(E)\Big|\ge \delta\Big\}\leq
C\, e^{-c\delta\sqrt{N\eta^*}}
\label{ncont}
\ee
holds  for all $\eta^*$ satisfying $K_\delta/N \leq \eta^* \leq 1$
and for all $N\ge 2$.
\end{itemize}
\end{theorem}

\medskip

As a corollary to this theorem, we can formulate a result on the eigenvectors:

\begin{corollary}\label{cor:linfty} Let $H$ be an $N\times N$ hermitian
Wigner matrix satisfying the condition {\bf C1)},
then the following hold:
\begin{itemize}
\item[(i)] For any
$\kappa>0$ and $K>0$ there exist constants  $C = C (\kappa,K)$
and $c=c(\kappa,K)$ such that for any interval $I\subset [-2+\kappa, 2-\kappa]$
of length $|I|\leq K/N$ we have
\be
\P \Bigg\{\exists \text{ $\bv$ with $H\bv=\mu\bv$,
$\| \bv \|=1$, $\mu \in I$ and } |v_1|
\ge \frac{M}{N^{1/2}} \Bigg\} \leq C e^{-c \sqrt{M}}\;
\label{efn}
\ee
for all $M \geq 0$ and $N\ge 2$.
\item[(ii)]
For any
$\kappa>0$, $K >0$, and  $2\leq p<\infty$
there exist $C= C(\kappa,K,p)$ and
$c=c(\kappa,K,p)>0$ such that for any interval $I\subset [-2+\kappa, 2-\kappa]$
of length $|I|=K/N$ we have
\be
\P \Bigg\{\exists \text{ $\bv$ with $H\bv=\mu\bv$,
$\| \bv \|=1$, $\mu \in I$ and } \|\bv\|_p
\ge M N^{\frac{1}{p}-\frac{1}{2}} \Bigg\} \leq C e^{-c\sqrt{M}}\;
\label{efn1}
\ee
for all $M \geq 0$ and all $N\ge 2$.

\item[(iii)] For any
$\kappa>0$ there exist $C=C(\kappa)$ and
$c=c(\kappa)$ such that
\be
\P \Bigg\{\exists \text{ $\bv$ with $H\bv=\mu\bv$,
$\| \bv \|=1$, $\mu \in [-2+\kappa , 2- \kappa]$ and }  \|\bv\|_\infty
\ge \frac{M}{N^{1/2}} \Bigg\} \leq C e^{-c \sqrt{M}}\;
\label{efn2}
\ee
for all $M \geq (\log N)^4$ and all $N\ge 2$.
\end{itemize}
\end{corollary}

The second main result is an upper bound on the tail distribution
of the eigenvalue gap; the proof is given in Section \ref{sec:tail}.

\begin{theorem}\label{thm:gap} Let $H$ be an $N\times N$ hermitian
Wigner matrix satisfying the
condition {\bf C1)}. Let $\kappa>0$ and
fix an energy $E\in [-2+\kappa, 2-\kappa]$.
Denote by $\lambda_\al$  the largest eigenvalue
below $E$ and assume that $\al\leq N-1$. Then  there are positive
constants $C$ and $c$, depending on $\kappa$,
such that
\be
  \P \Big(\lambda_{\al+1} -E\ge \frac{K}{N}, \; \al\leq N-1\Big)
\leq C\; e^{-c\sqrt{K}}
\label{gapdec}
\ee
for any $N\ge 1$ and any $K\ge 0$.
\end{theorem}

\bigskip

The third main result is the Wegner estimate for the
averaged density of states:

\begin{theorem}\label{thm:wegner}
Let $H$ be an $N\times N$ hermitian Wigner matrix satisfying the
condition {\bf C1)}  and condition {\bf C2)} with an exponent $a=5$
in \eqref{charfn}. Let $\kappa>0$, $0< \e\le 1$ and set $\eta= \e/N$.
Let $\cN_I$ be the number of eigenvalues in  $I:=[E-\eta/2, E+\eta/2]$.
Then there exists a constant $C$ such that
\be
 \P(\cN_I\ge 1)\leq \E \; \cN_I^2 \leq C \, \e
\label{n1}
\ee
uniformly for all $N\ge 10$, for all
$E\in [-2+\kappa,2-\kappa]$  and for all $\e \le 1$. In particular,
\be
\sup_{I\subset [-2+\kappa, 2-\kappa]}
 \sup_{N\ge 10} \; \E \Big[ \frac{\cN_I}{N|I|}\Big]\leq C \; ,
\label{eq:ds}
\ee
and therefore  the averaged density of
states, $\E \, \varrho_N(E)$, is an absolutely continuous measure
with a uniformly bounded density, i.e.
\be
\sup_{|E|\leq 2-\kappa} \sup_{N\ge 10} \E \, \varrho_N(E) \leq C\,
\label{ids}
\ee
(with a slight abuse of notations, $\E \, \varrho_N(E)$ denotes
the measure and its density as well).
The constant $C$ in \eqref{n1}, \eqref{eq:ds}
and \eqref{ids} depends only on $\kappa$ and on the constants
characterizing the distribution $\rd \nu$ via the conditions
{\bf C1)}--{\bf C2)}.
The estimates \eqref{n1}--\eqref{ids} hold for $N\le 10$ as well if, instead
of {\bf C1)} and {\bf C2)},
we assume that the density function,
$(const.)\exp(-\wt g)$,
of the diagonal matrix elements   satisfies
$\int_\bR |\wt g'(x)|\exp(-\wt g(x))\rd x <\infty$.
\end{theorem}

{\it Remark.} The proof of Theorem \ref{thm:wegner} also gives
a bound on the moments of the Stieltjes transform. By inspecting
the first step of the proof, we actually prove the stronger bound
\begin{equation}\label{eq:m-wegner}
\sup_{E \in [-2+\kappa,2-\kappa]}\; \sup_{N\ge 10}\; \sup_{0<N\eta\le 1}
 \, (N\eta)
\E \, |m(E+i\eta)|^2 \, \leq C(\kappa)
\end{equation}
and then we deduce (\ref{eq:ds}) from this estimate by using
$\cN^2_I/(N\eta) \le C\text{Im } m (E+i\eta) \le C|m(E+i\eta)|$.
The same argument used to prove (\ref{n1}) also gives bounds
on higher moments of $\cN_I$, of the form
\begin{equation}\label{eq:NIk} \sup_{E \in [-2+\kappa,2-\kappa]}
\;\sup_{N\ge 10}
\, \E \, \cN_I^k \leq C \eps , \qquad  I=\big[ E-\frac{\e}{2N},
E+\frac{\e}{2N}\big]
\end{equation}
uniformly in $\eps\le 1$ with a constant $C$ depending only on $k$ and
$\kappa$.
Both (\ref{eq:m-wegner}) and (\ref{eq:NIk}) extend to the
case $N \leq 10$, under the additional assumption
$\int_\bR |\wt g'(x)|\exp(-\wt g(x))\rd x <\infty$ on the density
$(\const .) \exp (-\wt g (x))$ of the diagonal elements.

\bigskip

Finally, the following theorem establishes an upper
bound on the level repulsion.

\begin{theorem}\label{thm:repul}
Let $H$ be an $N\times N$ hermitian Wigner matrix satisfying the
condition {\bf C1)}. Let $\kappa>0$, $\eps>0$ and
set $\eta = \e/N$. Let $\cN_I$ be
the number of eigenvalues in $I = [E-\eta/2, E+\eta/2]$.
Fix $k \in \N$, and assume that condition ${\bf C2)}$ holds with $a=k^2+5$.
Then, there exists a constant $C >0$, depending on $k$ and $\kappa$, such that
\be
\P ( \cN_I\ge k ) \leq C \; \e^{k^2}
\label{want2}
\ee
uniformly for all $\e >0$, for all  $N\ge N_0(k)$ and for
all $E\in [-2+\kappa, 2-\kappa]$.
\end{theorem}

\bigskip

All these estimates hold away from the spectral edges, i.e. for
$\kappa>0$, and the constants that are indicated
to depend on $\kappa$ blow up as $\kappa\to0$.
It is possible to obtain  the asymptotic dependence of the constants
on  $\kappa$
by following the proofs but the formulae are complicated.
In some simpler cases we computed these formulae,
see the remarks after Theorem   \ref{thm:sc-old}
and \ref{cor:linfty-old}.

\bigskip

The common starting point of all proofs is
Proposition \ref{prop:basic}.
Using the estimate $\text{Im} (a+bi)^{-1} \leq (a^2 + b^2)^{-1/2}$
on the right hand side of \eqref{basic},
we have
\be
 \cN_I
\leq C \eta\sum_{k=1}^N \frac{1}{(a_k^2 + b_k^2)^{1/2}}
\label{out}
\ee
with
$$
 a_k := \eta + \frac{1}{N}\sum_{\al=1}^{N-1}
\frac{\eta \xi_\al^{(k)}}{(\lambda_\al^{(k)}-E)^2 + \eta^2}, \qquad
b_k:= h_{kk} - E - \frac{1}{N}\sum_{\al=1}^{N-1}
\frac{ (\lambda_\al^{(k)}-E)
\xi_\al^{(k)}}{(\lambda_\al^{(k)}-E)^2 + \eta^2}\; ,
$$
where $a_k$ and $b_k$ are the imaginary and real
part, respectively, of the reciprocal of the summands in \eqref{basic}.
Theorems \ref{thm:sc1} and \ref{thm:gap} rely only on
the imaginary part, i.e.  $b_k$
in \eqref{out} will be neglected.
In the proofs of Theorems \ref{thm:wegner} and \ref{thm:repul}, however, we
make an essential use of $b_k$ as well. Since typically
$1/N \lesssim |\lambda_\al^{(k)}-E|$, we
note that $a_k^2$ is much smaller
than $b_k^2$ if $\eta\ll 1/N$
and this is the relevant regime for the Wegner estimate and
for the level repulsion.
Assuming a certain smoothness condition on the distribution $\rd \nu$
(condition {\bf C2)}), the distribution of
the variables $\xi_\al^{(k)}$ will also be smooth. Although $\xi_\al^{(k)}$
are not independent for different $\al$'s, they are sufficiently
decorrelated so that the distribution of $b_k$ inherits some
smoothness which will make the expectation value $(a_k^2 + b_k^2)^{-p/2}$
finite for certain $p>0$. This will give a bound on the $p$-th moment
on $\cN_I$ which will imply \eqref{n1} and \eqref{want2}.

\bigskip

\section{Semicircle law and delocalization on intermediate
scales}\label{sec:larger}

In this section we review the proof
of  the convergence to the semicircle law on intermediate energy
scales of the order $\eta \geq (\log N)^4/ N$. This convergence has already
been established in our previous work \cite{ESY2} but with a speed of
convergence uniform in $\eta$, for
$\eta \geq (\log N)^8/ N$. Our new estimate shows that the speed
of convergence becomes faster as $\eta$ increases (and we also
reduce the power of the logarithm from 8 to 4).
Moreover, we show that the results hold
under the condition {\bf C1)} only. Thus we obtain a stronger
version of our earlier results under weaker assumptions.

The following
result is an analogue of Theorem \ref{thm:sc1} for intermediate scales.
It states that the density of states regularized on any scale
$\eta\ge N^{-1}(\log N)^4$  converges to the Wigner
semicircle law in probability uniformly for all energies away from
the spectral edges.
Note, however, that the estimate for larger  scales is sufficiently
strong so that uniformity in the spectral parameter $z$ can be
obtained which is not expected for short scales.

\begin{theorem}\label{thm:sc-old}
Let $H$ be an $N\times N$ hermitian Wigner matrix satisfying the
condition {\bf C1)}. There exist a universal constant $c_1 >0$, and positive
constants  $C$, $c$, depending only on $\kappa$, such that
the following hold:
\begin{itemize}
\item[(i)]
Let the energy scale
$\eta$ be chosen such that $(\log N)^4/N\leq
\eta\leq 1$.
Then the Stieltjes transform $m_N(z)$ (see \eqref{Sti})
of the empirical eigenvalue distribution of the $N\times N$ Wigner matrix
satisfies
\be
\P \Big\{ \sup_{E \in [-2 + \kappa, 2 - \kappa]}
|m_N(E+i\eta)- m_{sc}(E+i\eta)| \ge \delta \Big\}
\leq C e^{-c \delta \sqrt{N\eta}}
\label{mcont-old}
\ee
for any $\delta \leq c_1 \kappa$ and  $N\ge 2$.

\item[(ii)] Let $\cN_{\eta^*}(E)=\cN_{I^*}$ denote
the number of eigenvalues in
the interval $I^*=[E-\eta^*/2, E+\eta^*/2]$. Then,
for any $\delta \leq c_1 \kappa$ there is a constant $K_\delta$ such that
\be
\P \Big\{ \sup_{|E|\leq 2-\kappa} \Big| \frac{\cN_{\eta^*}(E)}{N\eta^*}
- \varrho_{sc}(E)\Big|\ge \delta\Big\}\leq
C\, e^{-c\delta\sqrt{N\eta^*}}
\label{ncont-old}
\ee
holds  for all $\eta^*$ satisfying $K_\delta(\log N)^4/N\leq
\eta^* \leq c_2\kappa/K_\delta$
and for  $N\ge 2$. In other words,
on  the scale
$\eta^*$ with  $(\log N)^4 /N\ll \eta^*\ll 1$   we
have the convergence of the counting function as well.
\end{itemize}
\end{theorem}

{\it Remark}.  It is possible to follow the
dependence of the constants on the distance from the
spectral edges. For example, (\ref{mcont-old}) can be replaced by the bound
\[ \P \Big\{ \sup_{E \in [-2 + \kappa, 2 - \kappa]}
|m_N(E+i\eta)- m_{sc}(E+i\eta)| \ge \delta \Big\}
\leq C e^{-c \delta \sqrt{N\eta \kappa}}\]
for all $\delta \leq c_1 \kappa$, $(\log N)^4/N \leq \eta \leq 1$
and $N \geq 2$, for constants $C,c,c_1>0$ independent of $\kappa$.

\bigskip

{\it Proof.} This theorem  is proven exactly as
Theorem 1.1 in \cite{ESY2}  after replacing the key Lemma 2.1 of
\cite{ESY2} by the following Lemma \ref{lm:x-old}.
M. Ledoux has informed us that Lemma 2.1 of
\cite{ESY2} follows from a result of Hanson and Wright \cite{HW}.
We will reproduce his argument in the proof of Proposition \ref{prop:x}.
This requires only Proposition \ref{prop:HW} below, which is a mild extension
of the Hanson-Wright  theorem to the complex case.

\begin{lemma}\label{lm:x-old} Let $E\in [-2+\kappa, 2-\kappa]$.
Suppose that $\bv_\alpha$ and $\lambda_\alpha$ are eigenvectors
and eigenvalues
of an $N\times N$ random hermitian matrix $B$ with a
law satisfying the assumption of Theorem \ref{thm:sc-old}.
Let
$$
  X = \frac{1}{N} \sum_\al \frac{\xi_\al-1}{\lambda_\al-z}
$$
with $z=E+i\eta$, $\xi_\al = |\bb\cdot \bv_\al|^2$, where
the components of $\bb$ are i.i.d.
random variables, independent of $B$ and satisfying the condition {\bf C1)}.
Then there exists
a positive constant $c$ (depending on $\kappa$)
so that for every $\delta >0$, we have
\begin{equation}\label{eq:claim}
\P[ |X|\ge \delta] \leq 5\, e^{- c \min\{\delta \sqrt{N\eta}, \,
\delta^2N\eta\}}
\end{equation}
if $N\eta \geq (\log N)^2$ and $N$ is sufficiently large
(independently of $\delta$).
\end{lemma}
\noindent
For simplicity, we formulated the lemma for $N\times N$ matrices, but
it will be applied for the $(N-1)\times (N-1)$ minors of $H$.

\medskip

{\it Proof of Lemma \ref{lm:x-old}.}
Define the intervals $I_n= [E- 2^{n-1}\eta, E+2^{n-1}\eta]$
and let $M$ and $K_0$
be sufficiently large fixed numbers.
We have $[-K_0, K_0] \subset I_{n_0}$ with $n_0=
C\log (K_0/\eta)\le C\log (NK_0)$. Denote by $\Omega$ the event
\be
\Omega : = \Omega(M, K_0)= \Big\{ \max_n \frac{\cN_{I_n}}{N|I_n|}
\ge M \Big\}
\cup \{ \max_\al |\lambda_\al|\ge K_0\} \; ,
\label{omegadef}
\ee
where $\cN_{I_n}=|\{\al\; : \; \lambda_\al\in I_n\}|$ is the
number of eigenvalues in the interval $I_n$.
Therefore, if $\P_\bb$ denotes the probability w.r.t. the variable
$\bb$, we find
$$
  \P[ |X|\ge \delta] \leq \E \Big[ {\bf 1}_{\Omega^c}\cdot
 \P_\bb  \big[ |X|\ge \delta] \Big] + \P(\Omega) \; .
$$
We will prove below the following two propositions which complete
the proof of Lemma \ref{lm:x-old}. \qed

\begin{proposition}\label{prop:x} Assume condition {\bf C1)}.
Let $\Omega  = \Omega(M, K_0)$ be given by \eqref{omegadef}
and let $\eta\ge 1/N$. Then for sufficiently large and fixed $M, K_0$
there is
a positive  $c=c(M, K_0)$ such that for any $\delta>0$
$$
 \E \Big[ {\bf 1}_{\Omega^c}\cdot
 \P_\bb  \big[ |X|\ge \delta] \Big]  \leq
4\, e^{-c\min\{ \delta\sqrt{N\eta}, \, \delta^2N\eta\}} \; .
$$
\end{proposition}
\begin{proposition}\label{prop:omega} Assume condition {\bf C1)}.
Let $\eta$ be chosen such that $(\log N)^2/N\leq \eta\leq1$.
Then for sufficiently
large and fixed $M$ and $K_0$ there are positive constants $c,C$ such that
\be
\P\big[\Omega(M, K_0)\big] \leq C e^{-c\sqrt{MN\eta}} \; .
\label{Pom}
\ee
for all $N \geq 2$.
\end{proposition}

Both results are based on a theorem of Hanson and Wright \cite{HW},
extended to non-symmetric variables by Wright \cite{Wr}. The result
was formulated for real valued random variables. We do not
know if their theorems hold for general complex random variables,
but they hold true in two special cases,
namely when either the real and imaginary
parts of $b_j$ are i.i.d. or if the distribution of $b_j$ is
rotationally symmetric (see \eqref{hass}). We formulate
this easy extension of their result and we give the proof in the
Appendix.

\begin{proposition}\label{prop:HW}
Let $b_j$, $j=1,2,\ldots N$ be a sequence of complex i.i.d. random
variables with distribution $\rd\nu$
satisfying the Gaussian decay \eqref{x2} for some $\delta_0>0$.
Suppose that condition \eqref{hass} holds, i.e. either both the real
and imaginary parts are i.i.d. or the distribution $\rd\nu$ is
rotationally symmetric. Let $a_{jk}$, $j,k=1,2,\ldots N$ be
arbitrary complex numbers and let $\cA$ be the $N\times N$ matrix
with entries $\cA_{jk}:= |a_{jk}|$.   Define
$$
 X=\sum_{j,k=1}^N a_{jk} \big[ b_j\ov{b}_k -\E  b_j\ov{b}_k\big]\; .
$$
Then  there exists a constant $c>0$, depending only on $\delta_0, D$
from \eqref{x2},
such that for any $\delta>0$
$$
\P (|X|\ge \delta)\leq 4\exp\big( -c\min\{\delta/A,\; \delta^2/A^2\}\big)\;,
$$
where $A:= (\text{Tr}\, \cA\cA^t)^{1/2}=\big[\sum_{j,k} |a_{jk}|^2\big]^{1/2}$.
\end{proposition}

{\it Proof of Proposition \ref{prop:x}.}
Write $X$ in the form
$$
 X= \sum_{j,k=1}^N a_{jk} \big[ b_j\overline{b_k} -
\E b_j\overline{b_k}\big]\; ,
$$
where
$$
 a_{jk} = \frac{1}{N}\sum_\al
\frac{ \overline{ u_\al(j)} u_\al(k)}{\lambda_\al-z}\; .
$$
We have
$$
A^2: =  \sum_{j,k=1}^N |a_{jk}|^2  =\frac{1}{N^2}
\sum_\al\frac{1}{|\lambda_\al-z|^2} \; .
$$
On the set $\Omega^c$ we have
\be
\begin{split}
 A^2 =\frac{1}{N^2} \sum_{n= 0}^{n_0}
\sum_{ \lambda_\al\in I_n\setminus I_{n-1}}
\frac{1}{|\lambda_\al-z|^2}
\leq  \frac{1}{N^2}\sum_{n= 0}^{n_0}
\frac{\cN_{I_n}}{(2^n\eta)^2}
\leq \frac{2M}{N\eta} \;
\label{omc}
\end{split}
\ee
where we estimated the number of eigenvalues in $I_n\setminus I_{n-1}$
by $\cN_{I_n}$ and we set $I_{-1}:=\emptyset$.
Using Proposition \ref{prop:HW} we obtain
that
$$
   \E \Big[ {\bf 1}_{\Omega^c}\cdot
 \P_\bb  [ |X|\ge \delta] \Big]
\leq
4\exp\big( -c\min \{ \delta\sqrt{N\eta}, \, \delta^2 N\eta\}\big)\;
$$
where the constant $c$ depends on $M$ and
on $\delta_0, D$ from \eqref{x2}.
This completes the proof of Proposition \ref{prop:x}. \qed

\bigskip

{\it Remark.}  The same result can be proven by assuming
that the distribution $\rd\nu$ satisfies the logarithmic
Sobolev inequality, see Lemma 2.1 of \cite{ESY2};
the bound $\exp{(-c\delta(\log N)^2)}$ obtained there
can be easily improved to $C\exp{(-c\delta\sqrt{N\eta})}$
since the exceptional set $\Omega$ is defined differently.

\medskip

{\it Proof of Proposition \ref{prop:omega}.} Under  condition {\bf C1)},
we showed in
Lemma 7.4 of \cite{ESY} that
\be
 \P\{ \max_\al |\lambda_\al|\ge K_0\} \leq C e^{-cK_0^2N}
\label{tail}
\ee
for sufficiently large $K_0$.  To estimate  the large deviation
of $\cN_{I_n}$, we use the following  weaker version
of Theorem 2.1 of \cite{ESY}.

\begin{theorem}\label{thm:upp1}
Assume  condition {\bf C1)}. There exists constants $c ,C >0$, and $K_0$  such that
\be
\P \big\{ |m_N(x+iy)| \geq K \big\} \leq C e^{-c \sqrt{KNy}}
\label{mlde}
\ee
for all $x\in\bR$,
$y > (\log N)/N$,  $N\ge 2$ and  $K \ge K_0$.
In particular, if $I\subset \bR$ is an interval with
length $|I| \ge (\log N)/N$, we have
\be
  \P \big\{ \cN_I\ge KN|I|\} \leq C e^{-c\sqrt{KN|I|}}\; .
\label{uppp}
\ee
\end{theorem}

\medskip

Combining \eqref{tail} and \eqref{uppp} and recalling $N\eta\ge (\log N)^2$,
we have
$$
 \P (\Omega) \leq C\log (NK_0)
e^{-c\sqrt{MN\eta}} + C e^{-cK_0^2N} \leq C e^{-\wt c\sqrt{MN\eta}}
$$
completing the proof of Proposition \ref{prop:omega}. \qed

\bigskip

{\it Proof of Theorem \ref{thm:upp1}.} The proof  of \eqref{uppp}
is the same as
the proof of Theorem 2.1 in \cite{ESY} but in the
estimate (2.20) at the end of the proof we use the following
lemma instead of Corollary 2.4 to Lemma 2.3 from \cite{ESY}:

\begin{lemma}\label{lm:bour}
Assume condition {\bf C1)}.
Let the components of the vector $\bb\in \bC^{N-1}$
be complex i.i.d. variables with a common distribution $\rd \nu$
and let $\xi_\al = |\bb \cdot \bv_\al|^2$, where $\{ \bv_\al\}_{\al\in
{\cal I}}$
is an orthonormal set in $\bC^{N-1}$. Then for
$\delta\leq 1/2$ there is a constant $c>0$
such that
\be
  \P\big\{ \sum_{\al\in {\cal I}}\xi_\al \leq \delta m\big\} \leq
e^{-c\sqrt{m}} \;
\label{lm:xii}
\ee
holds for any ${\cal I}$,
where $m=|{\cal I}|$ is the cardinality of the index set ${\cal I}$.
\end{lemma}

We remark that  a stronger bound of the form $e^{-cm}$
was proven in Lemma 2.3 \cite{ESY}  under the
condition that $\mbox{Hess} \; g $ is bounded and  in the special
case when $g(x,y)$ was in the form $g(x)+g(y)$.
An alternative proof under the condition that the
support of $\rd\nu$ is compact is due
to J. Bourgain and it is reproduced in the
Appendix of \cite{ESY2}.
Using the stronger $e^{-cm}$ bound in \eqref{lm:xii},
the bound in \eqref{uppp} can be improved to
$e^{-cKN|I|}$. Here we present a proof that gives the weaker
bound but it uses no additional assumption apart from {\bf C1)}
and \eqref{hass}.

\medskip

We also note that although the statement of Theorem 2.1 of \cite{ESY}
only gives an upper bound on the density $\cN_I / (N |I|)$
for an interval $I = [E-\eta/2, E+ \eta/2]$, in its proof,
this quantity is first estimated by
\[ \frac{\cN_I}{N|I|} \leq  C \text{Im } m_N (E+i\eta) \le
C | m_N (E+i\eta)|
%C \, \text{Im } \frac{1}{N} \sum_k \frac{1}{H_N -E -i \eta} (k,k)
\leq \frac{C}{N} \sum_k \left| \frac{1}{H_N - E - i\eta} (k,k)
\right| \, ,
\]
and then we controlled the absolute value of the diagonal
elements of the resolvent. Hence, effectively, the proof of
Theorem 2.1 of \cite{ESY} provides an upper bound for the
absolute value of the Stieltjes transform $m (z)$, not just
for its imaginary part.  This yields \eqref{mlde}
and it completes the proof of Theorem
\ref{thm:upp1}. \qed

\bigskip

{\it Proof of Lemma \ref{lm:bour}.}  Let
$$
 X:= \sum_{i, j=1}^N a_{ij} \big[ b_i\ov{b}_j - \E \, b_i\ov b_j\big],
\qquad \mbox{with}\qquad
  a_{ij}: = \sum_{\al\in {\cal I}} \ov{v}_\al(i)v_\al(j)\, .
$$
Notice that  $\sum_{\al\in {\cal I} }\xi_\al = X + | {\cal I}|=X+m$
since  $\E \,\xi_\al =1$. By  $\delta\le 1/2$ we therefore obtain
$$
 \P\big\{ \sum_{\al\in  {\cal I}}\xi_\al \leq \delta m\big\}
\leq \P\big\{ |X| \ge \frac{m}{2}\big\}\; .
$$
Since
$$
  A^2:= \sum_{i,j=1}^N |a_{ij}|^2 = \sum_{\al,\beta\in  {\cal I}}
\sum_{i,j=1}^N\ov{v}_\al(i)v_\al(j)v_\beta(i)\ov{v}_\beta(j) = m \; ,
$$
by Proposition \ref{prop:HW}, we obtain
$$
 \P\big\{ \sum_{\al\in  {\cal I}}\xi_\al \leq \delta m\big\}
\leq \P\big\{ |X| \ge \frac{m}{2}\big\}\;
\leq 4\exp\Big( -c\min\big\{
\frac{m}{2A}, \frac{m^2}{4A^2}
\big\} \Big)\leq  e^{-c\sqrt{m}}.
$$
for some $c>0$. \qed

\bigskip

Using Theorem \ref{thm:sc-old}, we can prove delocalization of the
eigenvectors of $H$. In Theorem 1.2 of \cite{ESY2} we proved that
$\| \bv \|_{\infty} \leq (\log N)^{9/2} / N^{1/2}$
holds for all eigenvectors with probability bigger than $1- e^{-c(\log N)^2}$.
The following theorem is a generalization of this result
using the stronger estimates from Theorem \ref{thm:sc-old}.

\begin{theorem}\label{cor:linfty-old} Let $H$ be an $N\times N$ hermitian
Wigner matrix satisfying the condition {\bf C1)}. For any
$\kappa>0$  there exist constants $C,c>0$, depending on $\kappa$
such that
\be \label{eq:old-ev}
\P \Bigg\{\exists \text{ $\bv$ with $H\bv=\mu\bv$,
$\| \bv \|=1$, $\mu \in [-2+\kappa, 2-\kappa]$ and } \| \bv
\|_\infty \ge \frac{M}{N^{1/2}} \Bigg\} \leq C e^{-c \sqrt{M}}\;
\ee
for all $M \geq (\log N)^4$ and all $N\ge 1$ large enough.
\end{theorem}

{\it Remark.} Analogously to Theorem \ref{thm:sc-old},
it is possible to follow the $\kappa$-dependence of
the bound (\ref{eq:old-ev}). One finds
\[ \P \Bigg\{\exists \text{ $\bv$ with $H\bv=\mu\bv$,
$\| \bv \|=1$, $\mu \in [-2+\kappa, 2-\kappa]$ and } \| \bv
\|_\infty \ge \frac{M}{N^{1/2}} \Bigg\} \leq C e^{-c \kappa \sqrt{M}}\;
\] for all $M \geq (\log N)^4$ and $N \geq 2$.

\bigskip

{\it Proof of Theorem \ref{cor:linfty-old}.}
Let $\eta^* = M/N$ and partition the interval $[-2+\kappa, 2-\kappa]$
into $n_1= O(1/\eta^*)\leq O(N)$
intervals $I_1, I_2, \ldots, I_{n_1}$
of length $\eta^*$.
As before, let $\cN_{I}=|\{ \beta\; : \; \mu_\beta\in I\}|$
denote the number of eigenvalues in $I$. Let
$$
 c_1 :=\varrho_{sc}(2-\kappa)= \min \big\{ \varrho_{sc}(E) \; : \; E\in
[-2+\kappa, 2-\kappa]\big\} >0 \; .
$$
By using \eqref{ncont-old}
in Theorem \ref{thm:sc-old} and the fact that $N\eta^*\ge (\log N)^4$,
we have
\be
\P \left\{\max_n \cN_{I_n} \leq \frac{c_1}{2}\, N\eta^*\right\} \leq CN
e^{- c \sqrt{N\eta^*}}
\leq C e^{- \wt{c} \sqrt{N\eta^*}} \; .
\label{pmax}
\ee
Suppose that $\mu
\in I_n$, and that $H\bv = \mu \bv$. Consider the decomposition
\be
\label{Hd-old} H = \begin{pmatrix} h & \ba^* \\
\ba & B
\end{pmatrix}
\ee
where  $\ba= (h_{1,2}, \dots h_{1,N})^*$  and  $B$ is the $(N-1)
\times (N-1)$ matrix obtained by removing the first row and first column
from $H$. Let  $\lambda_\al$ and $\bu_\al$ (for $\al=1,2,\ldots , N-1$)
denote the eigenvalues and the normalized eigenvectors  of $B$.
Similarly to \cite{ESY2}, from the eigenvalue equation $H \bv = \mu \bv$
and from \eqref{Hd-old}
we find for the first component of $\bv =(v_1, v_2, \ldots , v_N)$ that
%that
%\be
%h v_1 + \ba \cdot \bw = \mu v_1, \quad \text{and } \quad \ba v_1 + B
%\bw = \mu \bw
%\ee
%with $\bw= (v_2, \dots ,v_N)^t$. {F}rom these equations
%we obtain
%$ \bw = (\mu-B)^{-1} \ba v_1 $ and thus
%$$
%\|\bw\|^2= \bw\cdot \bw = |v_1|^2 \ba\cdot (\mu -B)^{-2} \ba
%$$
%Since $\|\bw\|^2 = 1 - |v_1|^2$, we obtain
\be\label{v1} |v_1|^2 =
\frac{1}{1+ \ba\cdot(\mu -B)^{-2} \ba} = \frac{1}{1 + \frac{1}{N}
\sum_{\alpha} \frac{\xi_{\alpha}}{(\mu - \lambda_{\alpha})^2}}
\leq \frac{4 N [\eta^*]^2}{\sum_{\lambda_\alpha \in I_n} \xi_{\alpha}} \, ,
\ee
where in the second equality we set
$\xi_{\alpha} = |\sqrt{N} \ba \cdot \bu_{\alpha}|^2$
and used
the spectral representation of $B$. We recall that the eigenvalues
of $H$, $\mu_1\leq \mu_2 \leq \ldots\leq \mu_N$, and the eigenvalues
of $B$ are interlaced: $\mu_1 \leq \lambda_1\leq \mu_2\leq \lambda_2 \leq
\ldots \leq \lambda_{N-1} \leq \mu_N$ and the inequalities are strict
with probability one (see Lemma 2.5 of \cite{ESY}).
This means that
there exist at least $\cN_{I_n}-1$ eigenvalues  of $B$
in $I_n$.
Therefore, using that the components of any eigenvector are identically
distributed, we have
\begin{equation}
\begin{split}\label{lon}
\P \Big( \exists &\text{ $\bv$ with $H\bv=\mu\bv$, $\| \bv \|=1$,
$\mu \in [-2+\kappa, 2-\kappa]$ and } \| \bv \|_\infty \ge
\frac{M}{N^{1/2}} \Big)  \\
&\leq N n_1 \, \sup_n \P \Big( \exists \text{ $\bv$ with
$H\bv=\mu\bv$, $\| \bv \|=1$, $\mu \in I_n$ and } |v_1|^2 \ge
\frac{M^2}{N} \Big)\\
&\leq C\,  N^2 \sup_n \P \left( \sum_{\lambda_\alpha \in I_n}
\xi_{\alpha} \leq 4 \right) \\
&\leq  C \, N^2 \sup_n \P \left( \sum_{\lambda_\alpha \in I_n}
\xi_{\alpha} \leq 4 \text{ and } \cN_{I_n} \geq
\frac{c_1}{2}\, N \eta^* \right) + C\,  N^2  \sup_n \, \P
\left(\cN_{I_n} \leq \frac{c_1}{2}\, N \eta^* \right)
\\&\leq C  \,  N^2 e^{- \wt{c}\sqrt{ N\eta^*}} + C\,
N^2 e^{- \wt{c} \, \sqrt{N\eta^*}} \\
&\leq C e^{-c \sqrt{M}},
\end{split}
\end{equation}
for a sufficiently small $\wt{c} >0$ (we also
used that $N\eta^* = M \ge(\log N)^4$ in the last step).
Here we used Lemma \ref{lm:bour} to estimate the first
probability in the fourth line of \eqref{lon} and
\eqref{pmax} to estimate the second one.
\qed

\section{Upper bound for the density on short scales}\label{sec:upp}

Now we start our analysis on short scales $\eta\ge 1/N$.
As before, we always assume condition {\bf C1)}  in addition to \eqref{hass}.
We first show a large deviation upper bound on the number of eigenvalues
on short scales about a fixed energy $E$ away from the spectral
edges. This complements the estimate in Theorem \ref{thm:upp1}
that was valid for larger scales.

\begin{theorem}\label{thm:upp}
Let $\kappa>0$ and
fix an energy $E\in [-2+\kappa, 2-\kappa]$. Let $\nu,\beta$
be positive numbers such that $\nu+4\beta<1/2$. Let $\eta>0$
with $1\leq N\eta\leq CN^{\beta}$. Let
$$
\cN := \#\{ \al\; :\; \mu_\al \in I_\eta:=[E-\eta/2, E+\eta/2]\}\; .
$$
Then for any $2\leq M\leq \frac{CN^{\beta}}{N\eta}$, we have
\be
 \P \Big( \frac{\cN}{N\eta} \ge M\Big)
 \leq \Big(\frac{C}{M}\Big)^{\nu MN\eta}
\label{upperprob}
\ee
and
for any $1\leq p\leq CN^{\beta/2}$
\be
  \E \Big[\frac{\cN}{N\eta}\Big]^p  \leq
  C^p\Big(1+ \frac{p}{N\eta}\Big)^p \; .
\label{upperexp}
\ee
All constants depend on $\kappa$.
\end{theorem}

\bigskip

{\it Remark.} Similarly to the remark at
the end of the proof of Theorem \ref{thm:upp1},
we actually prove bounds on the absolute value
of the Stieltjes transform $m_N (z)$. So, instead of
(\ref{upperprob}), we actually show the stronger bounds
\be
\P \left( |m_N (E+i\eta)| \geq M \right) \leq
\left( \frac{C}{M} \right)^{\nu MN \eta}\, ,
\ee
and
\be
\E \, |m_N (E+i\eta)|^p \leq C^p \left( 1 + \frac{p}{N\eta} \right)^p.
\ee

\bigskip

{\it Proof of Theorem \ref{thm:upp}.}
It is sufficent to prove \eqref{upperprob}, since \eqref{upperexp}
easily follows from it and from \eqref{uppp}:
\be
\begin{split}
 \E \Big[\frac{\cN}{N\eta}\Big]^p \leq \; &
  2^p
+p\int_{2}^\infty M^{p-1} \P \Big( \frac{\cN}{N\eta} \ge M\Big)
\rd M\\
  \leq \; & 2^p +  p\int_2^{\Lambda} M^{p-1-\frac{\nu}{2} MN\eta}\;
\rd M
+   \int_{\Lambda}^\infty M^{p-1-c\sqrt{MN\eta}}\;\rd M\\
\leq \; & C^p\Big(1+ \frac{p}{N\eta}\Big)^p\;
\end{split}
\ee
with $\Lambda=CN^{\beta}/N\eta$ and with a sufficiently large constant $C$.
To prove \eqref{upperprob}, we use \eqref{basic} to obtain
\be
\frac{\cN_I}{N\eta} \leq C\sum_{k=1}^N \frac{1}{\eta +
\frac{1}{N}\sum_\al \frac{\eta\xi^{(k)}_\al} {(\lambda_\al^{(k)}-E)^2+\eta^2}}
\leq\frac{C}{N}\sum_{k=1}^N\frac{1}{\eta+
 \frac{Z^{(k)}}{N\eta }} \; ,
\label{keyestimate}
\ee
where we defined
\be
  Z^{(k)}:=Z^{(k)}(\eta)
= \sum_{\al: \lambda_\al^{(k)} \in I_\eta} \xi_\al^{(k)}\; .
\label{def:Z}
\ee
To estimate the large deviation of $Z^{(k)}$, we will later prove the
following lemma:

\begin{lemma}\label{lemma:lde} Let $\nu, \beta$ be positive numbers such
that $\nu+4\beta<1/2$. Then
for any
$\delta \ge N^{-2\nu}$
and $m\leq N^\beta$, we have
\be
\P\Big\{ \frac{1}{m}\sum_{\al=1}^m \xi_\al \le \delta\Big\}
\leq (C\delta)^{m}
\label{new}
\ee
with a constant $C$ depending on $\nu$ and $\beta$.
\end{lemma}
Note that the estimate in this
lemma is more precise than  \eqref{lm:xii}, but
the stronger estimate is valid only if $m$ is not
too large. In the proof we will use
information about the eigenfunctions obtained in Theorem
\ref{cor:linfty-old}.

\medskip

Let
$$
 \cN^{(k)}:=\cN^{(k)}_{I_\eta}=\#\{ \al: \lambda_\al^{(k)} \in I_\eta\}
$$
denote the number of
eigenvalues of the minor $B^{(k)}$ in the interval $I_\eta$
(see Section \ref{sec:not} for the definitions).
By the interlacing property of the eigenvalues,
$\cN\ge MN\eta$ implies $\cN^{(k)}\ge MN\eta-1 \ge
\frac{1}{2}MN\eta\ge N\eta$ for any $k$
(since $M\ge 2$ thus $MN\eta\ge 2$).
Therefore, from \eqref{keyestimate} we have for any $q\ge1$ that
\be
\begin{split}\label{Mest}
\P\Big( \frac{\cN}{N\eta}\ge M\Big)  \leq \; &
\P\Big(\frac{C}{N}\sum_k
\frac{{\bf 1}(\cN^{(k)}\ge \frac{1}{2}MN\eta)}{\frac{Z^{(k)}}{N\eta}
+ \eta} \ge M\Big) \\
\leq \; & \Big(\frac{C}{M}\Big)^q \E \Bigg[
\frac{{\bf 1}(\cN^{(1)}\ge\frac{1}{2}M N\eta)}{\frac{Z^{(1)}}{N\eta}
+ \eta}\Bigg]^q \\
\leq \; &\Big(\frac{C}{M}\Big)^q
\int_0^\infty \P \Bigg[ \cN^{(1)}\ge \frac{1}{2}MN\eta, \;
\frac{Z^{(1)}}{N\eta}
+ \eta\leq t^{-1/q}\Bigg]\rd t \\
\leq \; &\Big(\frac{C}{M}\Big)^q +\Big(\frac{C}{M}\Big)^q
\int_1^{(1/\eta)^q} \P\Bigg( \sum_{\al=1}^{MN\eta/2} \xi_\al^{(1)}
\leq N\eta  t^{-1/q}\Bigg)\rd t\\
\leq \; &\Big(\frac{C}{M}\Big)^q +\Big(\frac{C}{M}\Big)^q
\int_1^{(1/\eta)^q}
\big[C\max \{t^{-1/q}, N^{-2\nu}\}\big]^{\frac{1}{2}MN\eta} \rd t\\
\leq \; & \Big(\frac{C}{M}\Big)^{\nu MN\eta}
\end{split}
\ee
if we use Lemma \ref{lemma:lde} (noticing
that $\frac{1}{2}MN\eta\leq N^\beta$) and
we choose $q=\nu MN\eta$ in the last line
(we use that $N\eta\ge 1$).
$\Box$

\bigskip
{\it Proof of Lemma \ref{lemma:lde}.} We will present the
proof under the first condition in \eqref{hass}; the proof under the second
condition is analogous.
With the notation $\bb=\sqrt{N}\ba$, the components of $\bb$
can thus be written as $b_j = x_j+iy_j$ where $x_j$, $y_j$ are i.i.d.
random variables with expectation zero and variance 1/2.
Similarly we decompose the eigenvectors
into real and imaginary parts, i.e. we write $\bu_\al= \bv_\al+ i\bw_\al$
and we have
$$
\xi_\al = |\bb \cdot \bu_\al|^2 = \Big( \sum_{j=1}^N
(x_jv_\al(j)+ y_jw_\al(j))
\Big)^2 +\Big( \sum_{j=1}^N(x_jw_\al(j)- y_jv_\al(j))  \Big)^2 \; .
$$
The probability and expectation w.r.t. $\bb$ are denoted by $\P_\bb$
and $\E_\bb$.
We define the event
$$
\Omega:=\Big\{ \| \bu_\al\|_\infty \leq  CN^{2\beta-1/2}(\log N)^4\; :\;
\al=1,2,\ldots, m\Big\} \; ,
$$
where $\bu_\al$ are the eigenvectors of $B=B^{(1)}$.
Note that $\Omega$ is independent of the vector $\bb = \sqrt{N}
\ba^{(1)}$, thus
$$
\P\{\sum_{\al=1}^m \xi_\al \leq m\delta\} \leq \P (\Omega^c)
+ \E\Big[ {\bf 1}(\Omega)\P_\bb \big(\sum_{\al=1}^m \xi_\al \leq m\delta\big)
\Big]\;.
$$
By Theorem \ref{cor:linfty-old},
$$
 \P(\Omega^c)\leq e^{-cN^\beta (\log N)^2}\le (C\delta)^m\; .
$$
On the event $\Omega$, the probability
$\P_\bb \big(\sum_{\al=1}^m \xi_\al \leq m\delta\big)$
will be estimated as follows, where we introduced
$t:=\delta^{-1}\leq N^{2\nu}$:
\be
\begin{split}
\P_\bb\Big\{ \sum_{\al=1}^m \xi_\al \le m\delta\Big\}\leq \; &
e^{m}\E_\bb e^{-t\sum_{\al=1}^m \xi_\al} \\
=\; & e^{m}\E_\bb \prod_{\al =1}^me^{-t|\bb\cdot \bu_\al|^2}\\
=\; & e^{m}\E_\bb \prod_{\al=1}^m
\int_{\bR^2} \frac{\rd\tau_\al\rd s_\al}{\pi}
e^{-i\sqrt{t}\big[ \tau_\al \sum_j(x_jv_\al(j)+ y_jw_\al(j))
+ s_\al \sum_j(x_jw_\al(j)- y_jv_\al(j))  \big] - \tau_\al^2/4 - s_\al^2/4}\\
 =\; & e^{m}  \int_{\bR^{2m}} \prod_{\al=1}^m e^{- \frac{1}{4}
(\tau_\al^2+ s_\al^2)}
\frac{\rd\tau_\al\rd s_\al}{\pi}
  \\
&\quad \times
 \prod_{j=1}^N \E_{x_j}\E_{y_j}
 e^{-i\sqrt{t} \sum_\al \Big[ x_j (\tau_\al v_\al(j) + s_\al w_\al(j))
- y_j(s_\al v_\al(j) - \tau_\al w_\al(j))\Big]}\\
 \leq\; &  e^{m}   \int_{\bR^{2m}} \prod_{\al=1}^m \, e^{- \frac{1}{4}
(\tau_\al^2+s_\al^2)} {\bf 1}\big( |\tau_\al|+|s_\al|\le N^{\beta/2}
\log N\big)
\; \frac{\rd\tau_\al\rd s_\al}{\pi}
 \\
& \quad\times
 \prod_{j=1}^N \Bigg(1 - \frac{t}{8}\Big[ \big( \sum_\al
 (\tau_\al v_\al(j) + s_\al w_\al(j))\big)^2
+ \big( \sum_\al (s_\al v_\al(j) - \tau_\al w_\al(j))\big)^2\Big] \Bigg)\\
&  +
m(Ce)^me^{-cN^\beta(\log N)^2} + CNe^{-\delta_0N^{\beta}(\log N)^2}\;
.\label{pb}
\end{split}
\ee
The last two terms come from the Gaussian tail of the restriction
$|\tau_\al|, |s_\al| \leq N^{\beta/2}\log N$ for all $\al$,
and from the probability of the event  $\max_j |x_j|+ |y_j| \ge N^{\beta/2}
\log N$.
In estimate \eqref{pb} we have used that $\Big|\E\big[ e^{iY} - 1- iY +
\frac{1}{2}Y^2\big]\Big| \leq \E |Y^3|$, thus
for any real random variable $Y$
with $\E\, Y=0$ and $|Y|\leq \frac{1}{4}$ we have
$$
|\E \, e^{iY}| \leq
1- \frac{1}{2} \E\; Y^2 + \E |Y^3| \leq 1-\frac{1}{4} \E\;
Y^2 \; .
$$
We applied this to
$$
  Y =Y_j=
-\sqrt{t}\sum_{\al=1}^m \Big[ x_j (\tau_\al v_\al(j) + s_\al w_\al(j))
+ y_j(s_\al v_\al(j) - \tau_\al w_\al(j))\Big]
$$
with
$$
 \E \, Y^2_j = \frac{t}{2}\Big[ \big( \sum_\al
 (\tau_\al v_\al(j) + s_\al w_\al(j))\big)^2
+ \big( \sum_\al (s_\al v_\al(j) - \tau_\al w_\al(j))\big)^2\Big]
$$
and we also used that on the event $\max_j |x_j|+ |y_j| \le N^{\beta/2}\log N$
we have
\be
\begin{split}
%\Big|\sqrt{t}\sum_{\al=1}^m \Big[ x_j (\tau_\al v_\al(j) + s_\al w_\al(j))
%  + y_j(s_\al v_\al(j) - \tau_\al w_\al(j))\Big]\Big|
|Y_j| \leq \;  C m\sqrt{t}\,N^{3\beta-1/2}(\log N)^{6}
\leq \;  C N^{\nu+4\beta-1/2} (\log N)^{6} \leq \frac{1}{4}
% o( tN^{-1})\sum_\al (\tau_\al^2+s_\al^2)\;
\label{third}
\end{split}
\ee
on the event $\Omega$ and
in the regime where $|\tau_\al|, |s_\al| \leq N^{\beta/2}\log N$
for all $\al$.

Reexponentiating $1-\frac{1}{4} \E \, Y^2_j \leq \exp(- \frac{1}{8} \E Y^2_j)$
and using
that by the orthogonality of $\bu_\al$ we have,
$$
\sum_{j=1}^N \Big[\big( \sum_{\al=1}^m
 (\tau_\al v_\al(j) + s_\al w_\al(j))\big)^2
+ \big( \sum_{\al=1}^m (s_\al v_\al(j) - \tau_\al w_\al(j))\big)^2\Big]
=\sum_{\al=1}^m (\tau_\al^2 +s_\al^2) \;
$$
and we obtain (with $t=\delta^{-1}$)
\be
\begin{split}
\P_\bb\Big\{ \sum_{\al=1}^m \xi_\al \le m\delta\Big\}
\leq \; &
e^{m}
\int_{\bR^{2m}} \prod_{\al=1}^m e^{- \frac{1}{4}(\tau_\al^2 + s_\al^2) }
\frac{\rd\tau_\al\rd s_\al}{2\pi}
 e^{ - \frac{t}{16}\sum_\al (\tau_\al^2 +s_\al^2) }+
Ce^{-cN^\beta(\log N)^2} \\
\leq \; &
\Big(\frac{C}{1+t}\Big)^m + Ce^{-cN^\beta(\log N)^2} \\
\leq \; & (C\delta)^m \; .\qquad \Box
\end{split}
\ee

\bigskip

\section{Proof of the semicircle law on short scales}\label{sec:sc}

In this section we prove the semicircle law on the shortest
possible scale $\eta\ge O(1/N)$.

\bigskip

{\it Proof of Theorem \ref{thm:sc1}.}
We will prove only \eqref{sc:new}, the proof of \eqref{ncont}
can be obtained from \eqref{sc:new} exactly as in Corollary 4.2 of
\cite{ESY}.
We can assume that $\eta\leq (\log N)^4/N$, since the regime
$\eta\ge (\log N)^4/N$ has been covered in Theorem \ref{thm:sc-old}.
At the expenses of increasing the constant $C$ on the 
r.h.s. of (\ref{sc:new}), we can also assume that $N$ is sufficiently large.
 The constants in this proof depend on $\kappa$ (in addition
to  $\delta_0, D$ from \eqref{x2}) and
we will not follow their precise dependence. Set $z=E+i\eta$.
For $k=1,2, \ldots, N$ define the random variables
\be
X_k(z)=X_k: =\ba^{(k)}\cdot \frac{1}{B^{(k)}-z} \ba^{(k)}
- \E_k\; \ba^{(k)} \cdot \frac{1}{B^{(k)}-z} \ba^{(k)}
= \frac{1}{N}\sum_{\al=1}^{N-1} \frac{\xi_\al^{(k)} -1}
 {\lambda_\al^{(k)}-z} \; ,
\label{def:X}
\ee
where we used that $\E_k \xi_\al^{(k)}=\| \bu_\al^{(k)}\|^2=1$
and we recall that $\E_k$ denotes the expectation w.r.t. the random
vector $\ba^{(k)}$ (see Section \ref{sec:not} for notation).
We note that
$$
\E_k\; \ba^{(k)} \cdot \frac{1}{B^{(k)}-z} \ba^{(k)}
= \frac{1}{N}\sum_\al \frac{1}{\lambda_\al^{(k)}-z}
 = \Big(1-\frac{1}{N}\Big) m^{(k)} \;.
$$
It follows from \eqref{Sti} and \eqref{mm} that
\be\label{recur}
 m =  \frac{1}{N}\sum_{k=1}^N \frac{1}{ h_{kk} -z -
 \big(1-\frac{1}{N}\big)m^{(k)} - X_k} \; .
\ee

We use that
$$
   \Big|  m -  \Big(1-\frac{1}{N}\Big)m^{(k)}\Big|
=\Big| \int \frac{\rd F(x)}{x-z}
-  \Big(1-\frac{1}{N}\Big)\int \frac{\rd F^{(k)}(x)}{x-z}\Big|
= \frac{1}{N}\Big| \int \frac{NF(x)-(N-1)F^{(k)}(x)}{(x-z)^2} \rd
x\Big|.
$$
and we recall that the eigenvalues of $H$ and $B^{(k)}$ are interlaced,
\be
\mu_1\leq \lambda_1^{(k)}\leq \mu_2 \leq
\lambda_2^{(k)} \leq \ldots \leq \lambda_{N-1}^{(k)}
\leq \mu_N \, ,
\label{interlace}
\ee
(see e.g. Lemma 2.5 of \cite{ESY}), therefore
we have $\max_x|NF(x)-(N-1)F^{(k)}(x)|\leq 1$.  Thus
\be
 \Big|  m -  \Big(1-\frac{1}{N}\Big)m^{(k)}\Big|
\leq \frac{1}{N} \int \frac{\rd x}{|x-z|^2}
=  \frac{\pi}{N\eta}\, .
\label{mmm}
\ee

Let $M\ge 2$ be sufficiently large and fixed. Fix $E\in[-2+\kappa, 2-\kappa]$
away from the spectral edge. Assume for the moment only that $1/N\leq
\eta\le 1$.
Define $I_n = [E- 2^{n-1} \eta, E+ 2^{n-1}\eta]$,
and let $K_0$ be a sufficiently  large  fixed number.
For some constant $C=C(K_0)$
we have $[-K_0, K_0] \subset \bigcup_{ n = 0}^{C\log N}  I_n$.
Denote by $\Omega$ the event
\be\label{def:Omega}
\Omega : = \Big\{ \max_{n\leq C\log N}\; \frac{\cN_{I_n}}{N|I_n|}
\ge M\Big\}
\cup \{ \max_\al |\lambda_\al|\ge K_0\} \; .
\ee
Let $n_0$ be the largest non-negative
integer such that $2^{n_0}N\eta\leq (\log N)^4$, recall
that we assumed $N\eta\leq (\log N)^4$.
Similarly to the proof of Proposition  \ref{prop:omega},
by using \eqref{upperprob} with, say, $\nu=1/4$,
for short scales and \eqref{uppp}
for larger scales, we get
\be
\begin{split}
 \P(\Omega)\leq  & \; e^{-cN}+\sum_{n=0}^{n_0}
\Big(\frac{C}{M}
\Big)^{2^{n-3}MN\eta} +\sum_{n=n_0+1}^{C\log N}
e^{-c \sqrt{2^nMN\eta}} \\
\leq &\;
e^{-cN}+ \Big(\frac{C}{M}\Big)^{cMN\eta} + e^{-c\sqrt{MN\eta}} \leq 3\,
e^{-c\sqrt{N\eta}}
\label{omegaest}
\end{split}
\ee
with some $c>0$
(first term coming from the probability of $\max_\al |\lambda_\al|\ge K_0$).

{F}rom now on, we additionally assume that
$K/N\leq \eta \leq (\log N)^4/N$.
For $n\leq n_0$ define
$z_n= E+ i\eta_n $ with $\eta_n= 2^n\eta$, i.e. $z=z_0$
and $2^n\eta \leq (\log N)^4/N$ for all $n\leq n_0$.
We have  from \eqref{recur}
\be
\begin{split}\label{mplusm}
  m(z_n)
= \; &\frac{1}{N}\sum_{k=1}^N \frac{1}{-m(z_n) -z_n +\delta_k}\\
= \; & \frac{1}{-m(z_n) - z_n} -  \frac{1}{N} \sum_{k=1}^N
  \frac{1}{-m(z_n) - z_n}\; \frac{\delta_k}{h_{kk} - z_n - \frac{1}{N}
 \sum_{\al=1}^{N-1} \frac{\xi_\al^{(k)}} {\lambda_\al^{(k)}-z_n} }\; ,
\end{split}
\ee
where
$$
 \delta_k =\delta_k(z_n): = h_{kk} + m(z_n) - \Big(1-\frac{1}{N}\Big)
m^{(k)}(z_n) - X_k(z_n).
$$
Recall  $c_0=\pi \varrho_{sc}(E)$, thus $\mbox{Im}\; m_{sc}(z)=c_0+ O(\eta)$.
Define the event
$$
  \Xi_n: =\{ \mbox{Im}\; m(z_n) \ge c_0/10\}\; .
$$

On the event $\Xi_n$, by using \eqref{mmm} and that $N\eta\ge K= 300/c_0$,
we have $\mbox{Im}\; m^{(k)}(z_n)\ge c_0/20$ for any $k$.
Thus, on the event $\Xi_n\cap \Omega^c$ and
for any positive integer $r$, we have
\be
\begin{split}
  \frac{c_0}{20}\leq \; & \frac{1}{N}\sum_\al
\frac{\eta_n}{(\lambda^{(k)}_\al - E)^2+\eta_n^2}\\
\leq\; & \frac{\cN^{(k)}_{I_{n+r}}}{N\eta_n}+
\frac{1}{N}\sum_{\ell = n+r+1}^{C\log N} \sum_{\al\; : \; \lambda_\al^{(k)}
\in I_\ell
\setminus I_{\ell-1}}
\frac{\eta_n}{(\lambda^{(k)}_\al - E)^2+\eta_n^2}\\
\leq\; & \frac{\cN^{(k)}_{I_{n+r}}}{N\eta_n}+
\frac{1}{N}\sum_{\ell =n+r+1}^{C\log N}
\frac{2^n\eta \cN_{I_\ell}^{(k)}}{ (2^{\ell-2}\eta)^2}\\
\leq\; & \frac{\cN^{(k)}_{I_{n+r}}}{N\eta_n}+
16\sum_{\ell= n+r+1}^{C\log N}
\frac{\cN_{I_\ell}+1}{N|I_\ell|}\frac{1}{2^{\ell-n}}\\
\leq\; & \frac{\cN^{(k)}_{I_{n+r}}}{N\eta_n}+
2^{5-r}M\; ,
\end{split}
\ee
where we used that from the interlacing property we
have $\cN^{(k)}_{I}\leq \cN_I+1$,
for any interval $I$.

Thus, on $\Xi_n\cap \Omega^c$, with
the choice $r= [\log_2(1280M/c_0)]+1$,  we have the lower bound
$$
 \cN^{(k)}_{I_{n+r}}\ge \gamma_n\quad \mbox{with} \quad
\gamma_n:= \frac{c_0}{40}N\eta_n
$$
for any $n\leq n_0$ and for any $k=1,2, \ldots N$. Hence
from \eqref{mplusm} and recalling the definition \eqref{def:Z}
we get, for any $p\ge 1$, that
\be
\begin{split}\label{mm1new}
\E \Big|  m(z_n) + \frac{1}{m(z_n) + z_n}\Big|^p{\bf 1}\big(\Xi_n\cap
\Omega^c\big)
  \leq \; & \E \Bigg[ \frac{10}{c_0}\;  \frac{1}{N} \sum_{k=1}^N
  \frac{|\delta_k|\cdot
{\bf 1}(\cN_{I_{n+r}}^{(k)}\ge \gamma_n)  }{ \eta_n + \frac{1}{N} \sum_\al
\frac{\eta_n\xi_\al^{(k)} }{ (\lambda_\al^{(k)}-E)^2+\eta^2_n}}
\Bigg]^p \\
\ \leq \; & \E \Bigg[    \frac{10}{c_0}\;
  \frac{|\delta_1|\cdot
{\bf 1}(\cN_{I_{n+r}}^{(1)}\ge\gamma_n)  }{ \eta_n +
\frac{1}{2^{2r}N\eta_n}Z^{(1)}(\eta_{n+r})
} \Bigg]^p\\
 \leq \; & 2^{2pr}C_1^p \big[ \E \;  |\delta_1|^{2p} \big]^{1/2}
\Bigg[ \E \Bigg|\frac{1}{ \eta_n + \gamma_n^{-1}
\sum_{\al =1}^{\gamma_n} \xi_\al^{(1)}}
\Bigg|^{2p}
\Bigg]^{1/2}
\end{split}
\ee
(with $C_1 = (const)c_0^{-2}$).
The second term can be  estimated similarly to \eqref{Mest}. For
any  $1\leq p\leq  c_0N\eta/300$ we have that
\be
\begin{split}
\E \Bigg|\frac{1}{ \eta_n + \gamma_n^{-1}
\sum_{\al =1}^{\gamma_n} \xi_\al^{(1)}}
\Bigg|^{2p} \leq \; &
  \int_0^{(1/\eta_n)^{2p}}\P \Big(\sum_{\al=1}^{\gamma_n/2}\xi^{(1)}_\al
 \leq \gamma_n t^{-1/2p}\Big)\rd t\\
\leq \; & 1 +\int_1^{(1/\eta)^{2p}}
 \big[ C\max\{t^{-1/2p}, N^{-2\nu}\}\big]^{\gamma_n/2}\rd t\\
\leq \; &  C_\nu
\end{split}
\ee
where we chose e.g. $\nu=1/3$ and used that $2p/\nu \leq \gamma_n
\leq C(\log N)^4$.

For the first term on the r.h.s. of \eqref{mm1new},
we use $\E |h_{kk}|^{2p} \leq C^pN^{-p}$ and
\eqref{mmm} to get
\be
 \E \;  |\delta_1|^{2p} \leq C^pN^{-p} + \Big(\frac{C}{N\eta_n}\Big)^{2p}
+ C^p\E|X_1(z_n)|^{2p}.
\label{delta1}
\ee
To estimate $\E|X_1(z_n)|^{2p}$, we will need the following
extension  of Lemma \ref{lm:x-old} to $\eta\ge O(1/N)$.

\begin{lemma}\label{lm:x} Let $E\in [-2+\kappa, 2-\kappa]$.
Suppose that $\bv_\alpha$ and $\lambda_\alpha$ are eigenvectors
and eigenvalues
of an $N\times N$ random matrix with a
law satisfying the assumption of Theorem \ref{thm:sc-old}.
Let
$$
  X = \frac{1}{N} \sum_\al \frac{\xi_\al-1}{\lambda_\al-z}
$$
with $z=E+i\eta$, $\xi_\al = |\bb\cdot \bv_\al|^2$, where
the components of $\bb$ are i.i.d.
random variables satisfying the condition {\bf C1)}.
Then there exist two positive constants $K$, $C$ and $c$
(depending on $\kappa$)
so that for every $0<\delta \leq 1$, we have
\begin{equation}\label{eq:Xbound}
\P[ |X|\ge \delta] \leq C\; e^{- c \, \min\{ \delta \sqrt{N\eta},\;
\delta^2N\eta\} }
\end{equation}
if $K\leq N\eta \leq(\log N)^4$.
\end{lemma}

{\it Proof of Lemma \ref{lm:x}.} We follow the proof of
Lemma \ref{lm:x-old} but with the redefined set $\Omega$
(see \eqref{def:Omega} instead of \eqref{omegadef}).
Using the improved bounds
from Theorem \ref{thm:upp} we have already proved in
\eqref{omegaest} that $\P(\Omega)\leq 3\, e^{-c\sqrt{N\eta}}$.
To estimate $\E\big[ {\bf 1}_{\Omega^c}\cdot \P_\bb (|X|\ge \delta)\big]$,
we follow the proof of Proposition \ref{prop:x}. The only
difference is that in \eqref{omc} the summation runs from $n=0$
to $n=C\log N$, but the estimate on the right hand side of \eqref{omc}
is still valid. This completes the proof of Lemma \ref{lm:x}. \qed

\bigskip

Given the bound \eqref{eq:Xbound}, we have
$$
\E|X_1(z_n)|^{2p} \leq  \frac{(Cp^2)^p}{(N\eta_n)^p}
$$
and, from \eqref{delta1}, we get
$$
\E \;  |\delta_1|^{2p} \leq  \frac{(Cp^2)^p}{(N\eta_n)^p}\; .
$$
Thus
\be
\E \Big|  m(z_n) + \frac{1}{m(z_n) + z_n}\Big|^p{\bf 1}\big(
 \Xi_n\cap\Omega^c\big)
  \leq \frac{(Cp)^p}{(N\eta_n)^{p/2}} \; .
\label{concn}
\ee
For any $\delta$, set the event
$$
 \Lambda_n(\delta)=\Lambda_n:= \Big\{ \Big|
m(z_n) + \frac{1}{m(z_n) + z_n}\Big|\ge \delta\Big\}
$$
then from \eqref{concn}
$$
 \P(\Lambda_n\cap \Omega^c)\leq \P(\Xi_n^c\cap\Omega^c)
+ \frac{(Cp)^p}{(N\eta_n\delta^2)^{p/2}} \; .
$$

We recall the  stability of the equation $m+(m+z)^{-1}=0$, i.e. that
there exists a universal constant $C$ such that 
\be
  \Big| m(z)+ \frac{1}{m(z)+z}\Big|\leq \delta
\quad \Longrightarrow \quad |m(z)-m_{sc}(z)|\leq C_\kappa\delta\; 
\label{stabb}
\ee
with $C_{\kappa} = C \kappa^{-1/2}$, for all $\delta >0$
 and all $z$ with $|\text{Re}\, z| \leq 2-\kappa$, and 
$0 \leq \text{Im z} \leq 1$. To prove (\ref{stabb}), 
we observe that, from $m_{sc} + (m_{sc} +z)^{-1} =0$, 
and $|m+ (m+z)^{-1}| \leq \delta$, it follows that
\[ \begin{split} 
\delta \geq \;& \left| m -m_{sc} + \frac{1}{m+z} - \frac{1}{m_{sc} +z} 
\right| =
\left| (m-m_{sc}) \left(1-\frac{1}{(m+z)(m_{sc}+z)}\right) \right|
 = \left| (m-m_{sc}) \left(1+\frac{m_{sc}}{m+z}\right)\right| \\ 
= \; & |m-m_{sc}| \frac{|m+m_{sc}+z|}{|m+z|} 
\end{split} \]
and thus
\[ 
|m-m_{sc}| \leq \frac{\delta|m+z|}{|m+m_{sc}+z|} \,. 
\]
Next we observe that there exists a universal constant 
$C$ such that $|m_{sc} (z)| \leq C$ for all $z$ with
 $|\text{Re}\, z| \leq 2$ and $0 \leq \text{Im} \, z \leq 1$.
 Therefore
\[ 
|m- m_{sc}| \leq \frac{\delta |m+z| \, 
{\bf 1} (|m+z| \geq 2C)}{|m+z|-C} + \frac{2C \delta}{|m+m_{sc} +z|} 
\leq 2\delta + \frac{2C \delta}{\text{Im} \, m_{sc} (z)}
\]
where we used the fact that 
$\text{Im} \, m (z)>0$ and $\text{Im} \, m_{sc} (z) >0$ 
for all $z$ with $\text{Im } z>0$. Since 
\begin{equation}\label{eq:infmsc}
 \inf_{|\text{Re} \, z| \leq 2-\kappa, |\text{Im} \, z| \leq 1} \, 
\text{Im} \, m_{sc} (z) \geq \frac{c_0}{2} \geq C \sqrt{\kappa} 
\end{equation} 
for a universal constant $C$ we obtain (\ref{stabb}) 
(recall that $c_0 = \pi \varrho_{sc} (E) \geq C \sqrt{\kappa}$ 
for all $|E| \leq 2-\kappa$).

\medskip

Choosing $\delta \leq c_0/10C_\kappa$ 
(which can certainly be satisfied if $\delta \leq c_1 \kappa$ 
for a universal constant $c_1$) and using again (\ref{eq:infmsc}),
 we also see that
\be
  \Big| m + \frac{1}{m+z}\Big|\leq \delta
\quad \Longrightarrow \quad \mbox{Im} \; m \ge 2c_0/5 \, .
\label{lowb}
\ee
We also know (e.g. from \cite{ESY})
$$
 \mbox{Im} \; m(z_{n}) \ge \frac{1}{2} \mbox{Im} \; m(z_{n+1}) \; .
$$
Thus,  on the event $\Xi_n^c$ we have $\mbox{Im}\;  m(z_{n+1})\le c_0/5$,
which by \eqref{lowb} implies that
$$
   \Xi_n^c \subset \Lambda_{n+1}
$$
assuming that $\delta \leq 4c_0/5C_\kappa$.

Thus, we get
$$
\P (\Lambda_n\cap \Omega^c)\leq \P(\Lambda_{n+1}\cap\Omega^c) +
\frac{(Cp)^p}{(2^n\eta N\delta^2)^{p/2}} \; .
$$
Iterating this inequality up to $n_0$, we obtain
\[
\P (\Lambda_n\cap \Omega^c)\leq \P(\Lambda_{n_0}\cap\Omega^c) +
\sum_{j=n}^{n_0} \frac{(Cp)^p}{(2^j \eta N\delta^2)^{p/2}} \, .
\]
Using the result from
\cite{ESY2} on the scale $\eta_{n_0}\sim (\log N)^4/N$, we get
$$
\P (\Lambda_n\cap \Omega^c)\leq\frac{(Cp)^p}{(\eta N\delta^2)^{p/2}}
+e^{-c(\log N)^2}
$$
for sufficiently large $N\ge N_0$.
Thus, combining this with \eqref{omegaest},
for sufficiently small $\delta$,  we have
$$
\P \Big( \Big| m(z_0) + \frac{1}{m(z_0)+ z_{0}}\Big|\ge \delta
\Big) \leq \frac{(Cp)^p}{(\eta N\delta^2)^{p/2} }+
C\, e^{-c\sqrt{N\eta}} +e^{-c(\log N)^2}
$$
Choosing $p=\min\{ 1, \; c\delta \sqrt{N\eta}\}$
with some small constant $c$ and
using the stability bound  \eqref{stabb}, we obtain
Theorem \ref{thm:sc1} for the remaining case of
$\eta\leq (\log N)^4/N$.  $\Box$.

\bigskip

{\it Proof of Corollary \ref{cor:linfty}.}  Part (i)
follows from \eqref{v1} and \eqref{lon} by noticing
that no $N^2$ entropy factor in \eqref{lon} is needed.
In estimating $\P (\cN_{I_n} \leq \frac{1}{2} c_1 N\eta^*)$ in \eqref{lon}
we infer to the semicircle law \eqref{ncont}
which now holds on the $O(1/N)$ scale.
Part (ii) follows from part (i) and from
$$
 \P (\| \bv \|_p^p \ge M^p N^{1-\frac{p}{2}} ) =
 \P \Big( \frac{1}{N}\sum_{j=1}^N |v_j|^p \ge \frac{M^p}{N^{p/2}} \Big)
\leq \Big(\frac{N^{p/2}}{M^p}\Big)^q \E |v_1|^{pq}
\leq C e^{-c\sqrt{M}} \; .
$$
with the choice of $q=c\sqrt{M}$ where $c=c(\kappa, K,p)>0$
is sufficiently small and $C=C(\kappa,K,p)$ is sufficiently large.
Here we used that from part (i) we have that for any $m\ge 1$
$$
 \E \, (N^{1/2}|v_1|)^m
\leq M_0^m + m\int_{M_0}^\infty t^{m-1} e^{-c\sqrt{t}}\rd t
\leq (Cm)^{2m}
$$
where $C=C(\kappa,K)$.

Part (iii) also follows
from part (i) after summing up
the estimate \eqref{efn} for all spectral intervals
and for all coordinates $v_j$ of $\bv$ by using that
the distribution of $v_j$ is independent of  $j$. \qed

\section{Proof of the tail of the gap distribution}\label{sec:tail}

{\it Proof of Theorem \ref{thm:gap}.}
First notice that for any $K_0(\kappa)$ it is sufficient to prove
the theorem for all $K\ge K_0(\kappa)$, by adjusting the prefactor
$C=C(\kappa)$ in \eqref{gapdec}.
Second, it is sufficient to consider the
case of sufficient large $N\ge N_0(\kappa)$. By increasing $K_0(\kappa)$
to ensure $K_0(\kappa)\ge N_0(\kappa)^2$ if necessary, we can estimate
\be
\P(\lambda_{\alpha+1}-E\ge K/N, \; \al\leq N-1)
\leq \P(\max_\beta \lambda_\beta \ge \sqrt{K}-2)
\label{gg}
\ee
for any $K\ge K_0$ and $N\leq N_0$.
We recall
part i) of Lemma 7.3
of \cite{ESY}, i.e. that there is a constant $c>0$ such that
\be
   \P \{ \max_\beta \lambda_\beta\ge L\} \leq e^{-cL^2N}
\label{extre}
\ee
for all $L\ge L_0$ sufficiently large (both $c$ and $L_0$ depend
on the constants in \eqref{x2}).  Thus the probability in
\eqref{gg} can be estimated by $C\exp(-c\sqrt{K})$.

Next we treat the case $K\ge CN$ with some
large constant $C$. Since
$\lambda_{\al+1}\ge E+ K/N$  implies $\max_\beta \lambda_\beta
\ge K/N-2\ge L_0$ for a sufficiently large $C$,
and using \eqref{extre}, we obtain much stronger
bound of the form $\exp(-cK^2N)$
for the tail probability of $\lambda_{\al+1}$. For the rest of
the proof we can thus assume that $K\leq CN$ and both $K$ and $N$ are
sufficiently large, depending on $\kappa$.

The event $\lambda_{\al+1}\ge E+K/N$ implies that
there is a gap of size $K/N$ about $E' = E+K/2N$.
Fix a sufficiently large $M$ (depending on $\kappa$) and let
$z'= E'+i\eta$, with $\eta=K/(NM^2)$
and denote
$$
 \cN_j = \#\{ \beta\; : \; 2^{j-1}K/N\leq
|\lambda_\beta-E'|\leq 2^{j}K/N\} \; , \quad j=0, 1,2, \ldots
$$
On the set where $\max_\al |\lambda_\al|\leq K_0$, with some
large constant $K_0$,
we can estimate
\be
\begin{split}
 \mbox{Im}\; m(z') = &
\frac{1}{N}\sum_{\beta=1}^{N-1} \frac{\eta}{(\lambda_\beta -E')^2
+\eta^2} \\
\leq & \frac{\eta}{N}\sum_{j=0}^{C\log N} \frac{\cN_j}{(2^{j-1}
K/N)^2}\; .
%\\ \leq & \;
%\frac{8}{M^2}\sum_{j=0}^{C\log N} 2^{-j} \frac{\cN_j }{2^{j+1} K}\; .
\end{split}\label{im}
\ee
Define
$$
\Omega : = \max_\al\{ |\lambda_\al|\leq K_0\}
\cup \bigcup_{j=0}^{C\log N} \{ \cN_j \leq 2^{j+1}KM\},
$$
with a sufficiently large $K_0$,
then, similarly to the estimate \eqref{omegaest}, and together with
$K\leq C N$, we get
$$
\P(\Omega^c)\le Ce^{-c\sqrt{K}} \; .
$$
Then, on the set $\Omega$, we have from \eqref{im}
\be
\mbox{Im}\; m(z')\leq
\frac{16}{M}\; .
\ee
For large $M$ this implies that $ |\mbox{Im}\; m(z')-
\mbox{Im}\; m_{sc}(z')|\ge \frac{1}{2}\mbox{Im}\; m_{sc}(z')=:c_0>0$
and from Theorem \ref{thm:sc1} we know that
$$
\P (|m(z')-m_{sc}(z')|\ge c_0) \leq e^{-c\sqrt{N\eta}} =e^{-c'\sqrt{K}}\; ,
$$
where the constants  depend on $\kappa$.
Thus, recalling that $\al$ was defined to be the
index of the largest eigenvalue below $E$, we have
$$
 \P(\lambda_{\alpha+1}-E\ge K/N, \; \al\leq N-1)\leq \P(\Omega^c)
 +  \P (|m(z')-m_{sc}(z')|\ge c_0) \le  Ce^{-c\sqrt{K}}\; .
$$
This proves  Theorem \ref{thm:gap}. $\Box$

\section{Proof of the Wegner estimate}\label{sec:wegner}

{\it Proof of Theorem \ref{thm:wegner}.} We can assume $\e< 1/2$.
{F}rom the basic formulae
\eqref{basic}, \eqref{out} and using the Schwarz inequality,
we obtain that
\be
\begin{split}
\E \; \cN_I^2 \leq &\; C(N\eta)^2\; \E \Bigg[ \im\;
\frac{1}{h - z- \frac{1}{N}\sum_{\al=1}^{N-1}
 \frac{\xi_\al}{\lambda_\al-z}} \Bigg]^2\\
 \leq &\;C\e^2\; \E\; \Bigg[ \Big( \eta+ \frac{1}{N}\sum_{\al=1}^{N-1}
\frac{\eta\xi_\al}{(\lambda_\al-E)^2+\eta^2} \Big)^2
+ \Big(h - E -\frac{1}{N}\sum_{\al=1}^{N-1}
\frac{(\lambda_\al-E)\xi_\al}{(\lambda_\al-E)^2+\eta^2} \Big)^2\Bigg]^{-1}\;,
\end{split}\label{basic1}
\ee
where $h = h_{11}$ and $\lambda_\al =\lambda_\al^{(1)}$,
i.e.
the eigenvalues of the minor $B=B^{(1)}$ obtained from $H$
by removing the first row and column, and $\xi_\al = \xi_\al^{(1)}=
|\bb \cdot \bu_\al^{(1)}|^2$ where $\bb \equiv (b_1, \dots ,b_{N-1}) :=
\sqrt{N}(h_{12},h_{13}, \ldots , h_{1N})$.

Introducing the notation
\[ d_{\alpha}: = \frac{N(\lambda_{\al} - E)}{N^2 (\lambda_{\al} - E)^2 + \e^2},
\qquad
 c_\al: =  \frac{\e}{N^2 (\lambda_{\al} - E)^2 + \e^2},
\]
we have
\be
\E \; \cN_I^2 \leq C\e^2\;
 \E\; \Bigg[ \Big( \sum_{\al=1}^{N-1} c_\al \xi_\al \Big)^2
+ \Big(h - E -\sum_{\al=1}^{N-1} d_\al \xi_\al\Big)^2\Bigg]^{-1}.
\label{n12}
\ee

Let $\gamma$ be defined so that
$$
  \lambda_{\gamma} - E = \min \Big\{
\lambda_\al - E \; : \; \lambda_\al - E
\ge \frac{\e}{N}\Big\},
$$
i.e. $\lambda_{\gamma}$ is the first eigenvalue above $E+\e/N$.
Thus $\lambda_{\gamma} \leq \lambda_{\gamma+1}
\leq \lambda_{\gamma+2}\leq \lambda_{\gamma+3}$ are
the first four eigenvalues above
$E+\e/N$. If there are no four eigenvalues above $E+\e/N$,
then we use the four consecutive eigenvalues below $E-\e/N$,
as it will be clear from the proof, what matters is only that the signs
of $d_{\gamma+j}$, $j=0,1,2,3$,  are identical.
At the end of the proof we will consider the exceptional case
when there are less than four $\lambda$-eigenvalues both above $E+\e/N$ and
below $E-\e/N$, i.e. in this case
all but at most six eigenvalues are in
$[E-\e/N, E+\e/N]$.

We define then
\be
 \Delta := N(\lambda_{\gamma+3} - E).
\label{def:DELTA}
\ee
Note that, by definition,
$$
\e
\leq N(\lambda_{\gamma}-E)\leq \ldots \leq N(\lambda_{\gamma+3} - E) =\Delta\;,
$$
in particular $d_{\gamma}\ge d_{\gamma+1}\ge d_{\gamma+2}\ge d_{\gamma+3}$
(since the function $x\to x/(x^2+\e^2)$  is decreasing for $x\ge \e$)
and $c_{\gamma}\ge c_{\gamma+1}\ge c_{\gamma+2}\ge c_{\gamma+3}$
thus
\be
 \min_{j=0,1,2,3}
d_{\gamma+j} = \frac{\Delta}{\Delta^2+\e^2}\ge \frac{1}{2\Delta},
\qquad \min_{j=0,1,2,3} c_{\gamma+j} \ge \frac{\e}{\Delta^2}\;.
\label{mind}
\ee
\medskip

Next, we discard, in the first term in the denominator of (\ref{n12}),
all contributions but the ones from $\al = \gamma,\gamma+1$. We find
\begin{equation*}
\begin{split}
\E \; \cN_I^2\leq & \; C\e^2
 \; \E \; \Bigg[ \Big( c_{\gamma}\xi_{\gamma} +
c_{\gamma+1}\xi_{\gamma+1}
\Big)^2
+ \, \Big( h- E - \sum_{\al=1}^{N-1} d_{\alpha} \xi_\al \Big)^2\Bigg]^{-1}.
\end{split}
\end{equation*}
Note that $c_\al$ and $d_\al$ depend on the minor $B$ and are independent
of the vector $\bb$, so we can first take the expectation value with respect
to $\bb$.
In  Lemma \ref{lm:Ebb} below we give a general
estimate for such expectation values. Applying
\eqref{eq:r>1}  from Lemma \ref{lm:Ebb}
with $r=p=2$, $\beta_1=\gamma+2$, $\beta_2=\gamma+3$, and using
the estimates \eqref{mind}, we have
\begin{equation*}
\begin{split}
\E\; \cN_I^2\leq & \; C \e \E\; \Delta^3\; .
\end{split}
\end{equation*}
To estimate
the tail probability of $\Delta$, we note that
for any $K\ge \e$, the event $\Delta \ge K$
means that there must be an interval
of size $(K-\e)/4N$  between
$E+\e/N$ and $E+K/N$
with no $\lambda$-eigenvalue. {F}rom Theorem \ref{thm:gap} we have
$$
   \P(\cN^\lambda_J =0)\leq C\, e^{-c\sqrt{N|J|}}
$$
for any interval $J$ with length $|J|\geq 1/N$. Thus
$$
   \P(\Delta\ge t)\leq C\, e^{-c\sqrt{t}}, \qquad t\ge 1.
$$
Therefore $\E \; \Delta^3$ is finite and thus $\E \; \cN_I^2 \leq C\e$
is proven. The other statements in Theorem \ref{thm:wegner}
are easy consequences of this estimate.

Finally, we have to consider the case, when all but at most six
$\lambda$-eigenvalues are within $[E-\e/N, E+\e/N]$.
For all these eigenvalues $\lambda_\al$
we have $\frac{1}{2}\e^{-1}\leq c_\al\leq \e^{-1}$.
If $N-1\ge 9$, then there are at least  three eigenvalues in
$[E-\e/N, E+\e/N]$, we
denote them by $\lambda_{\gamma_1}, \lambda_{\gamma_2}$,
and $\lambda_{\gamma_3}$.
Then we have from \eqref{n12} and from \eqref{eq:sigma0}
of Lemma \ref{lm:Ebb} below that
$$
 \E \; \cN_I^2 \leq \e^2 \E  \Big( c_{\gamma_1}\xi_{\gamma_1} +
c_{\gamma_2}\xi_{\gamma_2} + c_{\gamma_3}\xi_{\gamma_3} \Big)^{-2} \leq
C\e^4\; .
$$
This completes the proof for $N\ge 10$.

The case $N<10$ requires a different argument. Let $f$ be a smooth
cutoff function supported on $[-1,1]$, $0\leq f\leq 1$ and $f(x)\equiv 1$
for $|x|\leq 1/2$, and let $F(s)=\int_{-\infty}^s f(x) \rd x$ its
antiderivative, clearly $0\leq F(s)\leq 2$. Write
$$
 \E \, \cN_I^2 \leq N \sum_{\al=1}^N
\E^*\Big[ \E^{**} f\Big( \frac{\mu_\al-E}{\e/N}\Big)\Big]\; ,
$$
where $\E^*$ is the expectation with respect to the off-diagonal
matrix elements and $\E^{**}$ is the expectation with
respect to the diagonal elements $x_{ii}$, $i=1,2,\ldots N$.
Since $N$ is bounded, it is sufficient to show that the expectation
inside the square bracket is bounded by $C\e$.
Let $\bx = (x_{11}, x_{22}, \ldots, x_{NN})$ and viewing
$\mu_\al$ as a function of $\bx$, we have
\be
  \nabla_\bx\Big[ F\Big(  \frac{\mu_\al-E}{\e/N}\Big)\Big]
 = N\e^{-1} f\Big( \frac{\mu_\al-E}{\e/N}\Big)\nabla_\bx \mu_\al \; .
\label{dere}
\ee
Simple first
order perturbation shows that
$$
 \frac{\partial \mu_\al}{\partial x_{ii}} = \frac{2}{\sqrt N} |\bv_\al(i)|^2
\;
$$
where $\bv_\al$ is the eigenvector of $H$ belonging to $\mu_\al$.
Notice that the components of the gradient in \eqref{dere}
are nonnegative and
their sum is $2/\sqrt{N}$. Thus, summing up each component of \eqref{dere},
we get
\be
\begin{split}
 \E^{**} f\Big( \frac{\mu_\al-E}{\e}\Big)
 & = \frac{\e}{2\sqrt{N}} \sum_{i=1}^N
\int_{\bR^N} \Big[\prod_{j=1}^N \rd \wt\nu(x_{jj})\Big]
\frac{\partial}{\partial x_{ii}}
\Big[ F\Big(  \frac{\mu_\al-E}{\e/N}\Big)\Big]
\\
& = (\mbox{const.})\frac{\e \sqrt{N}}{2} \int_\bR
\rd x_{11}\; e^{-\wt g(x_{11})}
\frac{\partial}{\partial x_{11}} \Bigg[
\int_{\bR^{N-1}}
F\Big(  \frac{\mu_\al-E}{\e/N}\Big)
\prod_{j=2}^N \rd \wt\nu(x_{jj})
\Bigg] \\
&\leq C\e\sqrt{N}.
\end{split}
\ee
In the last step we used
integration by parts,
the boundedness of $F$, the fact that $\rd \wt\nu$
is a probability measure and
that $\int_\bR |\wt g'(x)|\exp (-\wt g(x))\rd x$
is finite. Thus we obtained the Wegner
estimate for the small values of $N$ as well. \qed

\bigskip

The proof actually shows the following stronger result that
will be needed in Section \ref{sec:level}.
As before, let $\mu$'s be the eigenvalues of an $N\times N$ Wigner
matrix, and let $\gamma = \gamma(N)$ defined as
$$
  \mu_{\gamma} - E = \min \Big\{
\mu_\al - E \; : \; \mu_\al - E
\ge \frac{\e}{N}\Big\}\; .
$$
For any
positive integer $d$, let
\be
 \Delta_{d}^{(\mu)} = N(\mu_{\gamma(N)+d-1}-E)
\label{def:deltamu}
\ee
i.e. the rescaled distance from $E$ to the $d$-th $\mu$-eigenvalue
above $E+\e/N$.
If there are no $d$  $\mu$-eigenvalues above
$E+\e/N$, then we use the eigenvalues below $E-\e/N$
to define  $\gamma =\gamma(N)$ as
$$
  \mu_{\gamma} - E = \max \Big\{
\mu_\al - E \; : \; \mu_\al - E
\le -\frac{\e}{N}\Big\}
$$
and
\be
 \Delta_{d}^{(\mu)} = N(E- \mu_{\gamma(N)-d+1}) \; .
\label{def:deltamu1}
\ee
To unify the notation, let us introduce the
symbol
\be
\Delta_d^{(\mu)} =\infty
\label{def:deltamu2}
\ee
for the extreme case,
when there are at most $d-1$ eigenvalues above $E+\e/N$
and at most $d-1$ eigenvalues below $E-\e/N$; in particular
in this case all but at most $2d-2$ eigenvalues are
between $E-\e/N$ and $E+\e/N$.

\begin{corollary}\label{cor:weg} With the notation above,
for any $d\ge 5$, $N\ge 10$ and $M\in \N$
there is a constant $C=C_{M,d}$ such that
\be
\E \Big[ {\bf 1}(\cN_I\ge 1) \cdot  \big[\Delta_d^{(\mu)}\big]^M\cdot
{\bf 1}( \Delta_{d}^{(\mu)} <\infty) \Big]
\leq C\,\e \; .
\label{genweg}
\ee
\end{corollary}

{\it Proof.} We proceed as in the  proof of Theorem
\ref{thm:wegner} above. By  ${\bf 1}(\cN_I\ge 1)\leq\cN_I^2$
and following the estimates \eqref{basic1}--\eqref{n12}, we have
$$
\E \Big[ {\bf 1}(\cN_I\ge 1) \cdot  \big[\Delta_d^{(\mu)}\big]^M\cdot
{\bf 1}( \Delta_{d}^{(\mu)} <\infty) \Big]\leq
 C\e^2\;
 \E\; \frac{\big[\Delta_d^{(\mu)}\big]^M\cdot
{\bf 1}(\Delta_{d}^{(\mu)} <\infty)}{ \Big( \sum_{\al=1}^{N-1}
c_\al \xi_\al \Big)^2
+ \Big(h - E -\sum_{\al=1}^{N-1} d_\al \xi_\al\Big)^2}\;.
$$

With the notation \eqref{def:deltamu}, the $\Delta$ in \eqref{def:DELTA}
is actually $\Delta = \Delta_4^{(\lambda)}$, where the superscript
indicates that it is  defined in the $\lambda$-spectrum.
By the interlacing property and by $d\ge 5$, we have
$$
   \Delta = \Delta_4^{(\lambda)}\leq \Delta_d^{(\mu)}
 \leq \Delta_{d+1}^{(\lambda)}  ,
$$
thus
$$
 \E \Big[ {\bf 1}(\cN_I\ge 1) \cdot  \big[\Delta_d^{(\mu)}\big]^M\cdot
{\bf 1}( \Delta_{d}^{(\mu)} <\infty) \Big]
\leq
\e^2\;
 \E\; \frac{\big[\Delta_{d+1}^{(\lambda)}\big]^M\cdot
{\bf 1}(\Delta_{d+1}^{(\lambda)} <\infty)}{
\Big( \sum_{\al=1}^{N-1} c_\al \xi_\al \Big)^2
+ \Big(h - E -\sum_{\al=1}^{N-1} d_\al \xi_\al\Big)^2} \; .
$$
Now we perform the expectation with respect to the $\bb$ variables
as before; the numerator is independent of $\bb$. We get
$$
 \E \Big[ {\bf 1}(\cN_I\ge 1) \cdot  \big[\Delta_d^{(\mu)}\big]^M\cdot
{\bf 1}( \Delta_{d}^{(\mu)} <\infty) \Big]
\leq C\e \;
\E \big[\Delta^{(\lambda)}_4\big]^3
\big[\Delta_{d+1}^{(\lambda)}\big]^M\cdot
{\bf 1}(\Delta_{d+1}^{(\lambda)} <\infty) \leq C\e
$$
since the tail distribution of any $\Delta_{d}^{(\lambda)}$ decays
faster than any polynomial. \qed

\begin{lemma}\label{lm:Ebb} Fix $p \in \N / \{ 0 \}$ and let $N\ge p+3$.
Let $\bu_1, \bu_2, \ldots , \bu_{N-1}$ be an arbitrary orthonormal
basis in $\bC^{N-1}$ and set $\xi_\al = |\bb\cdot \bu_\al|^2$, where
the components of $\bb$ are i.i.d complex variables with
distribution $\nu$ with density $h$ satisfying
the condition {\bf C2)} with an exponent $a=p+3$ in \eqref{charfn}.
Fix different  indices $\al_1, \dots ,\al_p, \beta_1, \beta_2
\in \{ 1, 2, \dots ,N-1 \}$. Assume that $c_j > 0$,
for $j =1 ,\dots ,p$. Let $d_{\alpha} \in \bR$ for all $1 \leq \al \leq N-1$
be arbitrary numbers
such that $d_{\beta_1}, d_{\beta_2} >0$. Then, for every $1 < r < p+1$,
there exists a constant $C_{r,p} < \infty$ such that
\begin{equation}\label{eq:r>1}
\E_{\bb} \,  \left[ \Big( \sum_{j=1}^p c_j \xi_{\al_j}
\Big)^2 + \Big(E - \sum_{\al=1}^{N-1} d_{\al} \xi_\al\Big)^2
\right]^{-\frac{r}{2}}
\leq \frac{C_{p,r}}{ (\prod_{j=1}^p c_j)^{\frac{r-1}{p}} \,
\min (d_{\beta_1}, d_{\beta_2})} \,.
\end{equation}
Moreover, for every $p \geq 3$, we also have the improved bound
\begin{equation}\label{eq:impr}
\E_{\bb} \,  \left[ \Big( \sum_{j=1}^p c_j \xi_{\al_j}
\Big)^2 + \Big(E - \sum_{\al=1}^{N-1} d_{\al} \xi_\al\Big)^2
\right]^{-\frac{p}{2}}
\leq \frac{C_{p}}{ (\prod_{j=1}^{p-2} c_j) \, \min (c_{p-1}, c_p) \,
\min (d_{\beta_1}, d_{\beta_2})} \,.
\end{equation}
for a constant $C_p$ depending only on $p$.

%For $p\ge2$, the constant in this inequality can be chosen
%$$
%    C_{p,r} =\frac{C^p}{(p!)^2(p-r+1)\om_{p+3}^{p+2}}.
%$$
%For the case $r=1$ and $p\ge 2$ we find
%\begin{equation}\label{eq:r1}
%\sup_E \E_{\bb} \, \left[ \Big( \sum_{j=1}^p c_j \xi_{\al_j}
% \Big)^2 +  (E - \sum_{\al=1}^{N-1} d_{\al} \xi_\al)^2 \right]^{-\frac{1}{2}}
%\leq 1 + C\; \frac{1+|\log \min c_j|}
%{ \min (d_{\beta_1}, d_{\beta_2})}
%\end{equation}

\medskip

Without the second term in the denominator, we have the following estimates:
For all $1\leq r < p$, there exists a constant $C_{p,r} < \infty$ such that
\begin{equation}\label{eq:sigma0}
\E_{\bb} \left[\; \sum_{j=1}^p c_j \xi_{\al_j}\right]^{-r}
\leq \frac{C_{p,r}}{(\min  c_j)^{r}}\; .
\end{equation}
%The constant in this inequality can be chosen
%$$
%    C_{p,r} =\frac{C^p}{(p!)^2(p-r)\om_{p+1}^p}.
%$$
%For $r = p-1$ (if $p \geq 2$) we also have, with
%$c_1 = \min  c_j$,
%\begin{equation}\label{eq:sigma0log}
%\E_{\bb} \left[\;\sum_{j=1}^p c_j \xi_{\al_j}\right]^{-(p-1)} \leq
% \frac{C_p \, |\log \, c_1|}{(\min_{j=2,\dots ,p} c_j)^{p-1}}
%\end{equation}
%for an appropriate constant $C_p$.
\end{lemma}
%[WE can remove \eqref{eq:r1} and \eqref{eq:sigma0log},
%the other two are the clean statements]

{\it Remark.} For (\ref{eq:sigma0}),
it is enough to assume that
\begin{equation} |\wh{h} (t,s)| \leq \frac{1}{(1  + \om_{p+1}( t^2+s^2))^{p+1}}
\end{equation}
instead of both conditions in \eqref{charfn} with exponent $a=p+3$.

\bigskip

{\it Proof.} To prove (\ref{eq:r>1}), we perform a
change variables from
$\bb =(b_1, \ldots , b_{N-1})$ to $\bz =(z_1, \ldots z_{N-1})$
by introducing
$$
 \bz = U^* \bb
$$
where $U$ is the unitary matrix with columns $(\bu_1, \ldots, \bu_{N-1})$.
Notice that the Jacobian is one, thus
\be
\begin{split}
\text{I} := \; & \E_{\bb} \, \left[ \left( \sum_{j=1}^p c_j
\xi_{\al_j} \right)^2 +  \Big(E - \sum_{\al=1}^{N-1} d_{\al}
\xi_\al\Big)^2 \right]^{-r/2}
= \int \frac{\rd\mu(\bz)}{[ P(\bz)]^{r/2}}
\end{split}\label{eq:bb}
\ee
with
$$
  \rd\mu(\bz) : = e^{-\Phi(\bz)}\prod_{\al=1}^{N-1} \rd z_\al
\rd \overline{z}_\al, \qquad
\Phi(\bz):= \sum_{\ell=1}^{N-1}
g\left( \re \, (U\bz)_{\ell}, \;
\im \, (U\bz)_\ell  \right)
$$
and
$$
 P(\bz) : = \Big( \sum_{j=1}^p c_{j} |z_{\al_j}|^2 \Big)^2 +
\Big(E - \sum_{\al=1}^{N-1} d_{\al} |z_\al|^2 \Big)^2.
$$
We define, for $t \in \bR$,
\be
\label{def:F}
F (t) := \int_{-\infty}^t \rd s \; \left( \Big( \sum_{j=1}^p c_j
|z_{\al_j}|^2 \Big)^2 + s^2 \right)^{-r/2} \;.
\ee
Note that, for every $r >1$, there exists a constant $C_r <\infty$, such that
\be\label{eq:Fbd}
0 \leq F (t) \leq \,
\frac{C_r}{\left( \sum_{j=1}^p c_{j} |z_{\al_j}|^2 \right)^{r-1}}
\ee
for every $t \in \bR$. For $j=1,2$, we have
\begin{equation*}
\begin{split}
z_{\beta_j} \frac{\rd}{\rd z_{\beta_j}} \; F
\left(  E- \sum_{\al=1}^{N-1} d_\al |z_{\al}|^2 \right)
=  - \frac{ d_{\beta_j} |z_{\beta_j}|^2 }{ [P(\bz)]^{r/2}}\; .
\end{split}
\end{equation*}
Introducing the first order differential operator
$$
  D : = z_{\beta_1} \frac{\rd}{\rd z_{\beta_1}} +
z_{\beta_2} \frac{\rd}{\rd z_{\beta_2}} \;,
$$
we find
\begin{equation}\label{eq:der1}
D  \left[ F   \Big(E- \sum_{\al=1}^{N-1} d_\al |z_{\al}|^2\Big) \right] =
- \frac{d_{\beta_1} |z_{\beta_1}|^2 + d_{\beta_2} |z_{\beta_2}|^2}{
[P(\bz)]^{r/2}}\; .
\end{equation}
{F}rom (\ref{eq:bb}), we get
\begin{equation}
\text{I} = \,   -  \int \, \rd\mu(\bz)
\frac{1}{d_{\beta_1} |z_{\beta_1}|^2 + d_{\beta_2} |z_{\beta_2}|^2}
\; D  \left[ F  \Big(E- \sum_{\al=1}^{N-1} d_\al |z_{\al}|^2\Big) \right] \,.
\end{equation}
Integrating by parts and using the fact that
$$
\frac{\rd}{\rd z_{\beta_1}} \; \frac{z_{\beta_1}}{d_{\beta_1}
|z_{\beta_1}|^2 + d_{\beta_2} |z_{\beta_2}|^2} + \frac{\rd}{\rd z_{\beta_2}}
\;\frac{z_{\beta_2}}{d_{\beta_1} |z_{\beta_1}|^2
+ d_{\beta_2} |z_{\beta_2}|^2}
= \frac{1}{d_{\beta_1} |z_{\beta_1}|^2 + d_{\beta_2} |z_{\beta_2}|^2} ,
$$
we find
\begin{equation}\label{eq:bb2}
\begin{split}
\text{I} = \; &  \int \, \rd\mu(\bz)
\frac{F \left(  E- \sum_{\al=1}^{N-1} d_\al |z_{\al}|^2 \right)}{d_{\beta_1}
|z_{\beta_1}|^2 + d_{\beta_2} |z_{\beta_2}|^2} \Big(1-
D \Phi(\bz)\Big).
\end{split}
\end{equation}
Clearly
$$
|D\Phi(\bz)|^2\leq (|z_{\beta_1}|^2+ |z_{\beta_2}|^2) \Bigg(
\Big|\frac{\partial \Phi(\bz)}{\partial z_{\beta_1}}\Big|^2+
\Big|\frac{\partial \Phi(\bz)}{\partial z_{\beta_2}}\Big|^2\Bigg).
$$
By a Schwarz inequality in  (\ref{eq:bb2}) and
using (\ref{eq:Fbd}), we have
\be\label{eq:A+B}
\text{I} \leq  C_r \;\frac{ A + B_1 +B_2}{
\min(d_{\beta_1}, d_{\beta_2})},
\ee
where
\be\label{eq:A,B}
\begin{split}
A:= &  \int \rd\mu(\bz)
\frac{1}{\left( \sum_{j=1}^p c_j |z_{\al_j}|^2 \right)^{r-1}}
\frac{1}{ |z_{\beta_1}|^2 +  |z_{\beta_2}|^2}
\\
B_k:= &  \int \rd\mu(\bz)
\frac{1}{\left( \sum_{j=1}^p c_j |z_{\al_j}|^2 \right)^{r-1}}
\Big|\frac{\partial \Phi(\bz)}{\partial z_{\beta_k}}\Big|^2.
\end{split}
\ee

The integral $\text{A}$ can be bounded as follows
\begin{equation}\label{eq:sumA}
\text{A} \leq  A_1 + A_2 + A_3
\ee
with
\be
\begin{split}
A_1:= \; & \int \rd\mu(\bz)
\frac{{\bf 1}\big(\sum_{j=1}^p  c_j |z_{\al_j}|^2\leq  \kappa \big) }
{\left( \sum_{j=1}^p c_j |z_{\al_j}|^2 \right)^{r-1}}
\\
A_2:= \; & \frac{1}{\kappa^{r-1}}\int \rd\mu(\bz)
\frac{1}{ |z_{\beta_1}|^2 +  |z_{\beta_2}|^2}\\
A_3:= \; &
\int \rd\mu(\bz)
\frac{{\bf 1}(|z_{\beta_1}|^2 +  |z_{\beta_2}|^2\leq 1) \cdot
{\bf 1} \left( \sum_{j=1}^p c_j |z_{\al_j}|^2 \leq \kappa \right)}
{\left( \sum_{j=1}^p c_j |z_{\al_j}|^2 \right)^{r-1}
(|z_{\beta_1}|^2 +  |z_{\beta_2}|^2)}
\end{split}
\end{equation}
for any $\kappa>0$.
We start with the estimate of $A_3$.
Decompose $z_{\al_j} = x_j + iy_j$ and $z_{\beta_j}= x_{p+j}+ iy_{p+j}$
into real and imaginary parts.
We define the function
\begin{equation}\label{eq:f}
 f(x_1, \ldots x_{p+2}, y_1, \ldots , y_{p+2})
: = \frac{ {\bf 1}\left(\sum_{j=1}^p c_j (x_j^2 + y_j^2)
\leq \kappa\right)\cdot
  {\bf 1}\left(\sum_{j=p+1}^{p+2} (x_j^2 + y_j^2)\leq 1\right)}
 {\Big[\sum_{j=1}^p c_j (x_j^2 + y_j^2)\Big]^{r-1}
\sum_{j=p+1}^{p+2} (x_j^2 + y_j^2)}
\end{equation}
on $\bR^{2p+4}$. Changing variables $c^{1/2}_j x_j \to x_j$,
$c_j^{1/2} y_j \to y_j$ and
using that $r < p+1$, it is simple to check that
$$
\| f \|_1 \leq
\frac{C_{r,p} \, \kappa^{p+1-r}}{\prod_{j=1}^p c_j}\, .
$$
Thus, recalling that $z_{\al}= (U^*\bb)_\al$ and since the indices
$\al_1, \ldots, \al_p$, $\beta_1, \beta_2$ are all distinct, we find,
by taking the Fourier transformation in the $x_1, \ldots x_{p+2}$,
$ y_1, \ldots , y_{p+2}$ variables, that
\be
\begin{split}\label{a3est}
 A_3 \leq & \; \|\wh{f} \|_{\infty}
\int_{\bR^{2p+4}} \prod_{j=1}^{p+2} \rd t_j \rd s_j
\; \Big| \E_\bb e^{-i\sum_{j=1}^p[ t_j \re (U^*\bb)_{\al_j}
+s_j \im (U^*\bb)_{\al_j}]-i\sum_{j=p+1}^{p+2}[ t_j \re (U^*\bb)_{\beta_j}
+s_j \im (U^*\bb)_{\beta_j}] }
\Big|
\\ \leq & \; \|f\|_1
\int_{\bR^{2p+4}} \prod_{j=1}^{p+2} \rd t_j \rd s_j
\Big| \E_\bb e^{- i [\re (U\bt')+\im (U\bs')]\cdot \re \bb
- i[\re (U\bt')-\im (U\bs')]\cdot \im \bb }\Big|
\\ \leq &\;\|f\|_1
 \int_{\bR^{2p+4}} \prod_{j=1}^{p+2} \rd t_j \rd s_j
\frac{1}{ \left( 1+ \om_{p+3}
\|U\bt'\|^2 + \om_{p+3}\|U\bs'\|^2\right)^{p+3}}
\\ \leq &\;\|f\|_1
\int_{\bR^{2p+4}} \prod_{j=1}^{p+2} \rd t_j \rd s_j
\frac{1}{ \left( 1+ \om_{p+3}
 \|\bt\|^2 + \om_{p+3}\|\bs\|^2\right)^{p+3}} \\  \leq \; &\frac{C_{r,p}
\, \kappa^{p+1-r}}{\prod_{j=1}^p c_j}
\end{split}
\ee
for an appropriate constant $C_p$. Here the components $t'_j$ of the vector
$\bt'\in \bR^{N-1}$ are defined to be all zero except $t'_{\al_j}: = t_j$,
$t'_{\beta_{p+1}}: = t_{p+1}$,
$t'_{\beta_{p+2}}: = t_{p+2}$; the vector $\bs'$ is defined similarly.
In the last but one step we used the bound \eqref{charfn} with exponent
$p+3$ for the Fourier transform of the distribution of $\bb$.
In the last step, we used that for the Euclidean norm
$\|U\bt'\|=\| \bt'\| = \| \bt \|$
with $\bt = (t_1, \ldots t_{p+2})$ and similarly
$\|U\bs'\|= \| \bs \|$ with $\bs = (s_1, \ldots, s_{p+2})$.
Using similar arguments to bound the terms $A_1$ and $A_2$, we conclude that
$$
A \leq C_{r,p} \left( \frac{1}{\kappa^{r-1}} +
\frac{\kappa^{p+1-r}}{\prod_{j=1}^p c_j} \right)
$$
for arbitrary $\kappa >0$. Optimizing over $\kappa$, we find
\be \label{Abound} A \leq
\frac{C_{r,p}}{\left( \prod_{j=1}^p c_j \right)^{\frac{r-1}{p}}} \, .
\ee

\medskip

To control the integrals $\text{B}_k$, $k=1,2$ in \eqref{eq:A+B},
we integrate by parts and we use that
$\beta_k\neq \al_j$:
$$
B_k = -\int \prod_{\al=1}^{N-1} \rd z_\al \rd \bar z_\al
\;
\left( \sum_{j=1}^p c_j |z_{\al_j}|^2 \right)^{-(r-1)}
\frac{\partial \Phi(\bz)}{\partial z_{\beta_k}}  \;
 \frac{\partial e^{-\Phi(\bz)}}{\partial \bar z_{\beta_k}}
=  \int \rd\mu(\bz)
\left( \sum_{j=1}^p c_j |z_{\al_j}|^2 \right)^{-(r-1)}
\frac{\partial^2\Phi(\bz)}{\partial z_{\beta_k}\partial\bar z_{\beta_k}}.
$$
Simple calculation shows that
$$
 \frac{\partial^2\Phi(\bz)}{\partial z_{\beta}\partial\bar z_{\beta}}
 = \frac{1}{4} \sum_{\ell=1}^{N-1} |\bu_{\beta}(\ell)|^2
\Delta g\left( \re \, (U\bz)_{\ell}, \;
\im \, (U\bz)_\ell  \right)
$$
thus
\be
B_k =  \frac{1}{4} \sum_\ell | \bu_{\beta_k}(\ell) |^2 \;
\E_\bb \Bigg[
\left( \sum_{j=1}^p c_j |(U^*\bb)_{\al_j}|^2 \right)^{-(r-1)}
\Delta g(\re \, b_\ell, \im\, b_\ell) \Bigg]\; .
\label{bk}
\ee
For each fixed $\ell$, the estimate of the expectation value is
identical to that of $A_1$ if the density function
$e^{-g}$ for $b_\ell$ is replaced with $e^{-g}\Delta g$
(and all other $b_m$, $m \neq \ell$,
are still distributed according to $e^{-g}$).
Although $e^{-g}\Delta g$ is not a probability density,
it is only the decay of its Fourier transform
that is relevant to proceed similarly to the estimate \eqref{a3est}.
Having obtained uniform bound on the expectation in \eqref{bk}, we can
perform the summation over $\ell$ and we obtain
$$
 B_k \leq \frac{C_{p,r}}{\left( \prod_{j=1}^p c_j
\right)^{\frac{r-1}{p}}} \, .
$$
Combining this with \eqref{Abound} and \eqref{eq:A+B}, we have proved
\eqref{eq:r>1}.

\medskip

To prove (\ref{eq:impr}), we proceed as before up to (\ref{eq:A+B}).
This time, however,
we  bound the term $\text{A}$ in (\ref{eq:A,B}), with $r = p$, by
$$
A \leq A_4  + A_5 + A_6 + A_7,
$$
where
\begin{equation}
\begin{split}
A_4 &= \int \rd \mu(\bz)
\frac{{\bf 1} \left( |z_{\al_{p-1}}|^2 + |z_{\al_p}|^2 \leq 1 \right)}{\left(
\sum_{j=1}^{p-2} c_j |z_{\al_j}|^2 +
\wt c \left( |z_{\al_{p-1}}|^2 + |z_{\al_p}|^2 \right) \right)^{p-1}} \\
A_5 &= \int \rd \mu(\bz)
\frac{1}{\left( \sum_{j=1}^{p-2} c_j |z_{\al_j}|^2 + \wt c \right)^{p-1}} \\
A_6 &= \int \rd \mu(\bz)
\frac{ {\bf 1} ( |z_{\beta_1}|^2 +  |z_{\beta_2}|^2 \leq 1)}{\left(
\sum_{j=1}^{p-2} c_j |z_{\al_j}|^2 +
\wt c \right)^{p-1}}
\frac{1}{ |z_{\beta_1}|^2 +  |z_{\beta_2}|^2} \\
A_7 &= \int \rd\mu(\bz)
\frac{{\bf 1} \left( |z_{\al_{p-1}}|^2 + |z_{\al_p}|^2 \leq 1 \right)
{\bf 1} ( |z_{\beta_1}|^2 +  |z_{\beta_2}|^2 \leq 1)}{\left(
\sum_{j=1}^{p-2} c_j |z_{\al_j}|^2 +
\wt c \left( |z_{\al_{p-1}}|^2 + |z_{\al_p}|^2 \right) \right)^{p-1}}
\frac{1}{ |z_{\beta_1}|^2 +  |z_{\beta_2}|^2}
\end{split}
\end{equation}
with $\wt c = \min (c_{p-1}, c_p)$. We consider first the term $A_7$.
We decompose
$z_{\al_j} = x_j + iy_j$ and $z_{\beta_j}= x_{p+j}+ iy_{p+j}$
into real and imaginary parts. We define the function
\begin{equation}\label{eq:f1}
f(x_1, \ldots x_{p+2}, y_1, \ldots , y_{p+2})
: = \frac{{\bf 1} \left( \sum_{j=p-1}^p x_j^2 + y_j^2 \leq 1 \right) {\bf 1}
\left( \sum_{j=p+1}^{p+2} (x_j^2 + y_j^2) \leq 1\right)}
 {\Big[ \sum_{j=1}^{p-2} c_j (x_j^2 + y_j^2) +
\wt c \sum_{j=p-1}^p (x_j^2 + y_j^2)
\Big]^{p-1} \sum_{j=p+1}^{p+2} (x_j^2 + y_j^2)}
\end{equation}
on $\bR^{2p+4}$. Changing variables $c^{1/2}_j x_j \to x_j$,
$c^{1/2} y_j \to y_j$ for
$j=1, \dots ,p-2$, and then letting $r = \sum_{j=1}^{p-2} (x_j^2 + y_j^2)$ and
$w = \sum_{j=p-1}^p (x_j^2 + y_j^2)$, we find that
\begin{equation}
\begin{split}
\| f \|_1 \leq \frac{C_p}{\prod_{j=1}^{p-2} c_j} \int_0^1
\rd w \, w \int_0^{\infty}
\rd r \frac{r^{p-3}}{(r + \wt c \, w)^{p-1}}
%\leq \frac{C_p}{\prod_{j=1}^{p-2} c_j}
% \int_0^1 \rd w \, w \int_0^{\infty} \frac{\rd r}{(r+\wt c \, w)^2} \leq
\leq \frac{C_p}{\wt c \, \prod_{j=1}^{p-2} c_j }
\end{split}
\end{equation}
for an appropriate constant $C_p$. Proceeding as in (\ref{a3est}), we conclude
that
$$
A_7 \leq \frac{C_p}{\wt c \, \prod_{j=1}^{p-2} c_j}.
$$
The terms $A_4, A_5,A_6$ can be controlled similarly. Hence
$$
A \leq \frac{C_p}{\wt c\, \prod_{j=1}^{p-2} c_j} \;.
$$
The bound for $B$ in (\ref{eq:A,B}) can be obtained analogously
as in the proof
of (\ref{eq:r>1}) (with the same modifications used for the term $A$).
The proof of (\ref{eq:sigma0}) is similar (but much simpler). \qed

\section{Proof of the level repulsion}\label{sec:level}

{\it Proof of Theorem \ref{thm:repul}.}
We can assume that $\e<1/2$ and that  $k\ge 2$,
the $k=1$ case was proven in Theorem \ref{thm:wegner}.
We recall the notation $\Delta^{(\mu)}_d$ from
\eqref{def:deltamu}--\eqref{def:deltamu2} and we split
$$
\P (\cN_I\ge k) \leq (I) + (II)
$$
with
\be
\begin{split}
  (I) := & \P (\cN_I\ge k, \Delta^{(\mu)}_d=\infty) \\
 (II) := &\P (\cN_I\ge k, \Delta^{(\mu)}_d<\infty) \;
\end{split}
\ee
for  some positive integer $d$.
{F}rom the basic
formula \eqref{basic} we have
\begin{equation}
\begin{split}
\cN_I
%\eta \, \sum_{j=1}^N \im \,
%\frac{1}{h_{jj} - E - i\eta - \frac{1}{N} \sum_{\al}
% \frac{\xi^{(j)}_\al}{\lambda_\al^{(j)} - e - i \eta}} \\
\leq \; &\frac{C\e}{N}
\sum_{j=1}^N \left[ \left( \eta + \frac{\eta}{N} \sum_{\al=1}^{N-1}
\frac{\xi^{(j)}_\al}{(\lambda_\al^{(j)} - E)^2 + \eta^2} \right)^2 +
\left( E -h_{jj} + \frac{1}{N} \sum_{\al=1}^{N-1}
\frac{(\lambda_\al^{(j)} - E) \,
\xi^{(j)}_\al}{(\lambda_\al^{(j)} -E)^2 + \eta^2}\right)^2 \right]^{-1/2} \; .
\end{split}
\end{equation}
We introduce the notations
$\xi_\al = \xi^{(1)}_\al$, $\lambda_\al = \lambda_{\al}^{(1)}$,
$h = h_{11}$, and
$$
 c_\al = \frac{\e}{N^2 (\lambda_\al - E)^2 + \e^2}, \qquad
d_\al = \frac{N(\lambda_{\al} - E)}{N^2 (\lambda_\al - E)^2 + \e^2}
$$
as before.
Using a moment inequality, we get
$$
(I) \leq C_k \e^{k^2}  \,
\E \; \frac{{\bf 1}(\Delta_{d}^{(\mu)} =\infty)}{
\left(   \sum_{\al=1}^{N-1} c_\al \, \xi_\al \right)^{k^2}} \; .
$$
This term represents the extreme case, when all
but at most $2d-2$ eigenvalues are in $[E-\e/N, E+\e/N]$.
Choosing $d=2k$ and  assuming
that $N\ge k^2+4k$, we see that for at least
$k^2+1$  different $\al$-indices we have $\lambda_\al\in[E-\e/N, E+\e/N]$,
i.e. $\frac{1}{2}\e^{-1}\leq c_\al\leq \e^{-1}$. Using
\eqref{eq:sigma0} with $r=k^2$, $p=k^2+1$, we get
\be
 (I) \leq C_k \e^{2k^2} \; .
\label{eq:I}
\ee

Now we turn to the estimate of (II) and we will consider the
following somewhat more general quantity:
$$
 I_N(M, k,\ell): = \E \Big[{\bf 1} (\cN^{\mu}_I \geq k) \cdot
\big[ \Delta_{\ell}^{(\mu)}\big]^M\cdot
{\bf 1}(\Delta_{2k-\ell+4}^{(\mu)} <\infty)\Big]
$$
for any $M\in\N$ and $4\leq\ell \leq k+2$. The index $N$
refers to the fact that the $\mu$'s are the eigenvalues of an $N\times N$
Wigner matrix. The superscript $\mu$ in $\cN^{\mu}_I$ indicates
that it counts the number of $\mu$-eigenvalues.
Since by definition $ \Delta_{\ell}^{(\mu)}\ge1$,
we know that $I_N(M, k, \ell)$ is monotone increasing in $M$. Moreover,
with the choice $M=0$, $\ell=4$ we have
\be
 (II)\leq I_N(0, k, 4) \; .
\label{III}
\ee
Since the existence of $k$ $\mu$-eigenvalues
in the interval $I$ implies that
$\cN_I^{(j)} \geq k-1$ for all $j=1, \dots ,N$ (where $\cN_{I}^{(j)}$
denotes the number of eigenvalues $\lambda^{(j)}_{\al} \in I$),
we obtain that
\begin{equation}
\begin{split}
I_N(M,k,\ell) \leq \; &
C_k \, \eps^{k+1} \,
\E\; \frac{{\bf 1} (\cN^{(1)}_I \geq k-1) \cdot
\big[ \Delta_{\ell}^{(\mu)}\big]^M\cdot
{\bf 1}(\Delta_{2k-\ell+4}^{(\mu)} <\infty)}{ \left[
\left(   \sum_{\al=1}^{N-1} c_\al \, \xi_\al \right)^2 +
\left( E -h +  \sum_{\al=1}^{N-1} d_\al \, \xi_\al \right)^2
\right]^{(k+1)/2}} \;.
\end{split}
\end{equation}
By the interlacing property we have
$\Delta_{2k-\ell+3}^{(\lambda)} \leq \Delta_{2k-\ell+4}^{(\mu)}$
and $ \Delta_{\ell}^{(\mu)} \leq \Delta_{\ell+1}^{(\lambda)}$, thus we have
\begin{equation}
\begin{split}
I_N(M,k,\ell) \leq \; &
C_k \, \eps^{k+1} \,
\E\; \frac{{\bf 1} (\cN^{(1)}_I \geq k-1) \cdot
\big[ \Delta_{\ell+1}^{(\lambda)}\big]^M\cdot
{\bf 1}(\Delta_{2k-\ell+3}^{(\lambda)} <\infty)}{ \left[
\left(   \sum_{\al=1}^{N-1} c_\al \, \xi_\al \right)^2 +
\left( E -h +  \sum_{\al=1}^{N-1} d_\al \, \xi_\al \right)^2
\right]^{(k+1)/2}}\; .
\end{split}
\end{equation}
We split this quantity into two terms:
$$
 I_N(M,k,\ell) \leq\left(\text{A} + \text{B} \right)
$$
with
\begin{equation}\label{eq:A,Brep}
\begin{split}
\text{A} &: = C_k \, \eps^{k+1} \, \E
\frac{{\bf 1} (\cN^{(1)}_I \geq k+2) \cdot
\big[ \Delta_{\ell+1}^{(\lambda)}\big]^M\cdot
{\bf 1}(\Delta_{2k-\ell+3}^{(\lambda)} <\infty)}{ \left[ \left(
\sum_{\al=1}^{N-1} c_\al \xi_\al \right)^2 +
\left( E -h +  \sum_{\al=1}^{N-1}
d_\al \, \xi_\al \right)^2 \right]^{(k+1)/2}} \\
\text{B} &:= C_k \eps^{k+1} \,
\E \frac{{\bf 1} (k-1 \leq \cN^{(1)}_I < k+2) \cdot
\big[ \Delta_{\ell+1}^{(\lambda)}\big]^M\cdot
{\bf 1}(\Delta_{2k-\ell+3}^{(\lambda)} <\infty)}{ \left[
\left(  \sum_{\al=1}^{N-1} c_\al \xi_\al \right)^2 +
\left( E -h +  \sum_{\al=1}^{N-1} d_\al \, \xi_\al \right)^2
\right]^{(k+1)/2}} \, .
\end{split}
\end{equation}
To control the first term, we denote by
$\lambda_{\al_1}, \dots ,\lambda_{\al_{k+2}}$
the first $k+2$  $\lambda$-eigenvalues in the set $I_{\eta}$.
Then $c_{\al_j} \geq \frac{1}{2}\e^{-1}$,
for all $j=1, \dots ,k+2$ and therefore, by (\ref{eq:sigma0}),
\begin{equation} \label{eq:Ares}
\text{A} \leq C_k \, \eps^{k+1}
\E \frac{{\bf 1} (\cN^{(1)}_I \geq k+2) \cdot
\big[ \Delta_{\ell+1}^{(\lambda)}\big]^M\cdot
{\bf 1}(\Delta_{2k-\ell+3}^{(\lambda)} <\infty)}{
\left( \e^{-1} \sum_{j=1}^{k+2}
\xi_{\al_j} \right)^{k+1}} \leq C_k \, \e^{2k+2} I_{N-1}(M, k-1, \ell+1)\;
\end{equation}
using that $\Delta^{(\lambda)}_m$ is monotone increasing in $m$.

To control the term $\text{B}$ in (\ref{eq:A,Brep}), we choose the indices
$\al_1, \dots , \al_{k-1}$ so that $\lambda_{\al_j} \in I_{\eta}$ for all
$j=1, \dots ,k-1$. Since we know that there are at most $k+1$ eigenvalues in
$I_{\eta}$, there must be, either on the right or on the left of $E$,
$\lambda$-eigenvalues at distances larger than $\e/N$ from $E$ if $N\ge k+8$.
Let us suppose, for example, that there are four such
eigenvalues on the right of $E$.
Then, we define the index $\al_k$  so that
$$
  \lambda_{\al_k} - E = \min \Big \{ \lambda_\al - E \; : \; \lambda_\al - E
  > \frac{\e}{N} \Big\}
$$
i.e. $\lambda_{\al_k}$ is the first eigenvalue above $E+\e/N$. Moreover,
let $\al_{k+1} = \al_k + 1$, $\beta_1 = \al_k + 2$
and $\beta_2 = \al_{k+1}+3$.
Recalling the notation \eqref{def:deltamu},
we set $\Delta:=\Delta^{(\lambda)}_4 = N(\lambda_{\beta_2} - E)$.
By definition
$$
\e\le N(\lambda_{\al_{k}} -E) \leq N (\lambda_{\al_{k+1}} - E)
\leq N (\lambda_{\beta_1} - E) \leq N (\lambda_{\beta_2} - E) = \Delta.
$$
and $\min (d_{\beta_1}, d_{\beta_2})\ge \frac{1}{2\Delta}$.
Therefore
\begin{equation}\label{eq:node}
\begin{split}
\text{B} \leq C_k \, \eps^{k+1} \, \E  \;\frac{{\bf 1}
\left(\cN^{(1)}_I \geq k-1\right)\cdot
\big[ \Delta_{\ell+1}^{(\lambda)}\big]^M\cdot
{\bf 1}(\Delta_{2k-\ell+3}^{(\lambda)} <\infty)}{
\left[ \left(\sum_{j=1}^{k-1}
\e^{-1} \xi_{\al_j}
+ \frac{\e}{\Delta^2} (\xi_{\al_k} + \xi_{\al_{k+1}})  \right)^2 +
\left( E -h + \sum_{\al=1}^{N-1} d_\al \, \xi_\al \right)^2
\right]^{(k+1)/2}} \; .
\end{split}
\end{equation}
{F}rom (\ref{eq:node}), we find
\begin{equation}
\begin{split}
\text{B} \leq \; & C_k \e^{k+1}\;\E_{\lambda,h} \Bigg\{
\Big[ {\bf 1} \left(\cN^{(1)}_I
\geq k-1\right)\cdot
\big[ \Delta_{\ell+1}^{(\lambda)}\big]^M\cdot
{\bf 1}(\Delta_{2k-\ell+3}^{(\lambda)} <\infty)\Big] \\
& \times \E_{\bb}  \left[ \left(\sum_{j=1}^{k-1} \e^{-1} \xi_{\al_j} +
\e \Delta^{-2} (\xi_{\al_k} + \xi_{\al_{k+1}}) \right)^2 +
\left( E -h + \sum_{\al=1}^{N-1} d_\al \, \xi_\al \right)^2
\right]^{-\frac{k+1}{2}}  \Bigg\}\; .
\end{split}
\end{equation}
Using (\ref{eq:impr}) from Lemma \ref{lm:Ebb}
(with $p=k+1$, $c_j =\e^{-1}$ for
all $j=1, \dots, k-1$, $c_k = c_{k+1} = \e \Delta^{-2}$),
it follows that
\begin{equation}
\begin{split}
\text{B} \leq & \; C_k \, \e^{2k-1}\, \E_{\lambda}
\Big[ {\bf 1} \left(\cN^{(1)}_I
\geq k-1\right) \big[\Delta_4^{(\lambda)}\big]^3\cdot
\big[ \Delta_{\ell+1}^{(\lambda)}\big]^M\cdot
{\bf 1}(\Delta_{2k-\ell+3}^{(\lambda)} <\infty)\Big]\\
\leq & \;  C_k \, \e^{2k-1}\,I_{N-1}(M+3, k-1,  \ell+1)\; ,
\end{split}
\end{equation}
where we used that $\min (d_{\beta_1}, d_{\beta_2}) \geq 1/2\Delta$
and that $\Delta_4^{(\lambda)} \leq  \Delta_{\ell+1}^{(\lambda)}$.

Together with (\ref{eq:Ares}) and the monotonicity
of $I_{N-1}$ in $M$,  we obtain that
$$
I_N(M, k, \ell) \leq C_k \e^{2k-1} I_{N-1}(M+3, k-1, \ell+1)\; .
$$
Iterating this inequality, we arrive at
$$
  I_N(M, k, \ell) \leq C_k \e^{k^2-1} I_{N-k+1}(M+3(k-1), 1, \ell +k-1)\;.
$$
Recalling \eqref{III}, we have
$$
(II)\leq I_N(0, k, 4) \leq
C_k \e^{k^2-1} I_{N-k+1}(3(k-1), 1, k+3) \; .
$$
Finally, $I_{N-k+1}(M, 1, d)$ was exactly the quantity
that has been estimated by $C\e$ for
any $M$ and $d\ge 5$ in
Corollary \ref{cor:weg} (replacing $N$ by $N-k+1$), thus we have
$$
 (II) \leq C_k\e^{k^2} \; .
$$
Together with  \eqref{eq:I}, this completes the proof of Theorem
\ref{thm:repul}.\qed

\appendix

\section{Proof of Proposition \ref{prop:HW}}

We first consider the case, when the real and imaginary parts of $b_j$ are
i.i.d. (first condition in \eqref{hass}). We
split $a_{jk}$ and $b_j$ into real and imaginary parts,
$a_{jk} = p_{jk} + i q_{jk}$, $b_j = x_j + iy_j$, and
form the vector $\bw = (x_1, \ldots x_N,  y_1, \ldots y_N)\in \bR^{2N}$
with i.i.d. components.
We write $X= X_1 + iX_2$ where
$X_1 = \bw\cdot \bP \bw - \E \bw\cdot \bP \bw$, $X_2= \bw \cdot \bQ \bw
-\E \bw \cdot \bQ \bw$ with
symmetric real $(2N)\times (2N)$
matrices $\bP$ and $\bQ$, written in a block-matrix form as
$$
 \bP = \frac{1}{2}\begin{pmatrix}
P+P^t & Q-Q^t\cr Q^t-Q & P+P^t \end{pmatrix}, \qquad
\bQ = \frac{1}{2}\begin{pmatrix}
Q+Q^t & P^t-P\cr P-P^t & Q+Q^t \end{pmatrix} \; ,
$$
where $P=(p_{jk})$ and $Q=(q_{jk})$.
We define $\cP$ to be the symmetric matrix
whose entries are the absolute values of the matrix entries of $\bP$:
$$
\cP =  \frac{1}{2}\begin{pmatrix}  P^\dagger & P^\# \cr
P^\# & P^\dagger \end{pmatrix}, \qquad  (P^\dagger)_{jk} = |p_{jk}+p_{kj}|,
\qquad ( P^\#)_{jk} =  |q_{kj}-q_{jk}| \; .
$$
Then
$$
 \mbox{Tr}\, \cP^2 = \frac{1}{2} \sum_{j, k}
\Big( |p_{jk}+p_{kj}|^2 + |q_{kj}-q_{jk}|^2\Big) \leq 2\sum_{j,k} \big[
p_{jk}^2 +q_{jk}^2\big] \; .
$$
We apply the non-symmetric version of
of the Hanson-Wright theorem \cite{Wr}
for $X_1$ and $X_2$ separately; note that the components
of $\bw$ are i.i.d. Together with the bound
$\|\cP\|\leq \sqrt{\text{Tr}\,
\cP^2}$ we have
$$
   \P ( |X_1|\ge \delta) \leq 2\exp{(-c \min\{ \delta/\sqrt{
\text{Tr}\, \cP^2},
\delta^2/\text{Tr}\, \cP^2\})}
$$
for some constant $c$ depending on $\delta_0$ and $D$ from \eqref{x2}.
Similar estimate holds for $X_2$, so we have
$$
   \P ( |X|\ge \delta) \leq 4\exp{(-c \min\{ \delta/A,
\delta^2/A^2\})}
$$
where  $A^2= \sum_{j,k} |a_{jk}|^2 =\sum_{j,k}\big[
|p_{jk}|^2 + |q_{jk}|^2\big]$.

In the second case in \eqref{hass}, when the distribution of the complex random
variable $b_j$ is rotationally
symmetric, we can directly extend  the proof \cite{HW}
(note that  \cite{HW} uses the notation $X_j$ for  $b_j$).
We first symmetrize the quadratic form $X$ by replacing $a_{jk}$
with $\frac{1}{2}[a_{jk} + \ov{a}_{kj}]$.
We then follow the proof in \cite{HW} and note that
the only change is that $Z$ used starting from Lemma 2 in \cite{HW} will be a
standard complex Gaussian random variable and instead of $Z^2$ or $Z^{2n}$
we consider $|Z|^2=Z\ov Z$ and $|Z|^{2n}$, and
similarly $X^{2n}$ is  replaced by $|X|^{2n}$, $n=1,2,\ldots$.
With these changes, Lemma 1--6 in \cite{HW} hold true for
the complex case as well. In the proof of the theorem, starting
on page 1082 of \cite{HW}, instead of
$\prod_i \E X_i^{\al_i}  (X_i^2 - \E X_i^2)^{\beta_i}$
the expansion will contain terms of
the form
$\prod_i \E X_i^{\al_i} \ov X_i^{\al_i'} (|X_i|^2 - \E |X_i|^2)^{\beta_i}$.
Due to the rotational symmetry of the distribution, these terms are all
zero (case (i) on page 1082 of \cite{HW}) unless $\al_i=\al_i'$ for all $i$.
In the latter case, the bound
$|\E |X_i|^{2\al_i}(|X_i|^2 - \E |X_i|^2)^{\beta_i}| \leq
\lambda^{2\al_i+2\beta_i} \E |Z_i|^{2\al_i}(|Z_i|^2 - 1)^{\beta_i}$
holds with a sufficiently large $\lambda$ (depending
on $\delta_0$ from \eqref{x2}) exactly as in case (ii) on
page 1082 of \cite{HW}. From now on
the proof is unchanged and we obtain
$$
\P \Big( \Big|
\sum_{jk} a_{jk}(b_j\ov{b}_k -\E b_j\ov{b}_k) \Big|\ge\delta\Big)\leq
2\exp\big(-c\min(\delta/A, \delta^2/A^2)\big)\; ,
$$
where $\sum_{jk} \big|\frac{1}{2}[a_{jk}+\ov{a}_{kj}]\big|^2$
was estimated by $A^2=\sum_{jk}|a_{jk}|^2$ from above. \qed

\end{document}